\documentclass[10pt,journal]{IEEEtran}
\IEEEoverridecommandlockouts

\usepackage{graphicx}
\usepackage{subfigure}
\usepackage{amsmath,amsthm,amsfonts,amssymb}
\usepackage{enumerate}
\usepackage{cite}
\usepackage{bm}
\usepackage{bbm}
\usepackage{url}
\usepackage{array}
\usepackage{color}
\usepackage{algorithm2e}

\theoremstyle{plain}

\newtheorem{rem}{Remark}

\usepackage{booktabs}

\allowdisplaybreaks[4]

\begin{document}
\title{AttentionCode: Ultra-Reliable Feedback Codes for Short-Packet  Communications}

\author{
Yulin~Shao,~\IEEEmembership{Member,~IEEE},
Emre Ozfatura,
Alberto Perotti,
Branislav Popovic,
Deniz G\"und\"uz,~\IEEEmembership{Fellow,~IEEE}
\thanks{Y. Shao, E. Ozfatura, and D. G\"und\"uz are with the Department of Electrical and Electronic Engineering, Imperial College London, London SW7 2AZ, U.K. (e-mail: \{y.shao, m.ozfatura, d.gunduz\}@imperial.ac.uk).

A. Perotti and B. Popovic are with the Radio Transmission Technology Lab, Huawei Technologies Sweden AB, Kista 164-94, Sweden (e-mail: \{alberto.perotti, branislav.popovic\}@huawei.com).}
}

\maketitle

\RestyleAlgo{ruled} 
\SetKwComment{Comment}{/* }{ */} 

\begin{abstract}
Ultra-reliable short-packet communication is a major challenge in future wireless networks with critical applications. To achieve ultra-reliable communications beyond $99.999\%$, this paper envisions a new interaction-based communication paradigm that exploits feedback from the receiver. We present AttentionCode, a new class of feedback codes leveraging deep learning (DL) technologies. The underpinnings of AttentionCode are three architectural innovations: AttentionNet, input restructuring, and adaptation to fading channels, accompanied by several training 
methods, including large-batch training, distributed learning, look-ahead optimizer, training-test signal-to-noise ratio (SNR) mismatch, and curriculum learning. The training methods can potentially be generalized to other wireless communication applications with machine learning.
Numerical experiments verify that AttentionCode establishes a new state of the art among all DL-based feedback codes in both additive white Gaussian noise (AWGN) channels and fading channels. In AWGN channels with noiseless feedback, for example, AttentionCode achieves a block error rate (BLER) of $10^{-7}$ when the forward channel SNR is $0$ dB for a block size of $50$ bits, demonstrating the potential of AttentionCode to provide ultra-reliable short-packet communications.
\end{abstract}


\section{Introduction}
The rolling out of the fifth generation (5G) new radio (NR) has empowered a wide range of emerging applications that require ultra-reliable short-packet communications, such as  autonomous vehicles, industrial automation and control, tactile Internet, augmented/virtual reality, etc.
The most demanding use cases of these applications have a reliability requirement of $10^{-9}$ block error rate (BLER) \cite{URLLC,5G}.
5G NR supports ultra-reliable communications with $10^{-5}$ BLER \cite{5G}, which cannot satisfy the most stringent reliability requirement. These applications typically have short packet lengths between $10$ to $50$ bytes \cite{URLLC}, rendering  ultra-reliable communications even more challenging as the performance of existing channel codes degrades significantly with the decrease in the block length \cite{code2016}.
Achieving ultra-reliable short-packet communications with a BLER lower than $10^{-5}$ is a long-term pursuit for the fifth generation (5G) and beyond.

In this paper, we envision a new interaction-based communication paradigm for future wireless networks, where we assume the availability of a feedback link from the receiver to the transmitter.
The information bits in the new communication paradigm are channel-coded based on successive feedback from the receiver, yielding ultra-reliability.
Modern wireless communication systems include feedback in one form or another \cite{love2008,kurka2020deepjscc,PNC}. For example, the feedback can be regarding the channel state information (CSI) in time-varying channels, or in the form of an automatic repeat request (ARQ).

The classical feedback channel model was introduced by Shannon in 1956 \cite{Shannon}, wherein a transmitter (node A) and a receiver (node B) interact with each other via a feedforward channel and a feedback channel. Both are assumed to be discrete-time additive memoryless Gaussian-noise channels.
In Shannon's original formulation, the feedback is passive, noiseless, and unit-time delayed. That is, node B feeds back the raw received symbol to node A with a unit-time delay. Node A perfectly observes what B receives in the last interaction and generates the next coded symbol accordingly. Shannon proved an important result that the feedforward channel capacity is not increased by feedback. Nevertheless, feedback can provide higher reliability at finite block length. A capacity-achieving coding scheme, now known as the Schalkwijk-Kailath (SK) scheme, was developed in \cite{SK1,SK2} to exploit noiseless feedback. It has been proven in \cite{SK1} and follow-up works \cite{follow1,follow2,Gallager} that the SK scheme can provide a dramatic reduction in the error probability for a given finite block length and code rate.

The SK scheme, however, is extremely sensitive to feedback noise -- its performance degrades significantly in the presence of arbitrarily small feedback noise \cite{Weissman2008,Love2011,Kim2007ISIT,ModuloSK}.
In \cite{Kim2007ISIT}, the authors proved that no linear code incorporating noisy passive feedback can achieve a positive rate of communication. 
They proposed a nonlinear coding scheme based on a three-phase detection and retransmission protocol to cope with noisy feedback, which was shown to be asymptotically optimal. On the other hand, the authors in \cite{ModuloSK} interpreted the SK scheme with noisy feedback as a joint source-and-channel coding (JSCC) problem with side information at node A. To exploit the side information, a simple and efficient modulo SK scheme was proposed by means of modulo arithmetic at node B. Provided that the signal-to-noise ratio (SNR) of the feedback channel sufficiently exceeds that of the feedforward channel, the modulo SK scheme exhibits close-to-optimal performance.

The advent of deep learning (DL) technologies have brought revolutionary advances in a variety of disciplines \cite{AlphaGo,Transformer,FLOAC,Significant2020}, motivating the design of feedback codes via DL \cite{DeepCode,DEFC,DRFC}. 
The essence of DL-based feedback code design is to model the communication system, including the iterative encoding, feedback, and decoding processes, as an autoencoder.
Feedback codes can then be discovered algorithmically by optimizing the autoencoder such that the reconstruction error at the receiver is minimized.
The pioneering work of DeepCode \cite{DeepCode}, for example, proposed a learning-based architecture with passive feedback to learn good feedback codes in additive white Gaussian noise (AWGN) channels, wherein both the encoder and decoder are designed to be deep neural networks (DNNs).
Given that linear codes cannot achieve a positive communication rate when the feedback link is noisy \cite{Kim2007ISIT}, DNNs, thanks to their ability to approximate highly nonlinear functions, have the potential to learn good feedback codes.

Compared with the human-crafted SK and modulo-SK schemes, DL-based feedback codes have two main advantages:
1) Free from precision issues and quantization errors. As detailed in Section \ref{sec:II}, the essence of the SK and modulo-SK schemes is transmitting a $2^K$-ary pulse-amplitude
modulation (PAM) constellation to the receiver for a bitstream of length $K$, and improving the estimate of the PAM constellation at the receiver by successive feedback and revision.
This iterative process, however, requires high-precision arithmetic and high-order quantization (more than $K$ bits) of electronic parts and components in communication systems.
In contrast, DL-based feedback codes do not suffer from this constraint.
2) Flexibility and generalizability. The SK and modulo-SK schemes are designed for unit-time delay feedback and AWGN channels with a specific pair of feedforward and feedback SNRs.
In an effort to design practical feedback codes, especially those that generalize well to fading channels, DL-based feedback codes are much-desired thanks to the flexibility and generalizability of DL.

Unlike prior works that relied exclusively on the recurrent neural network (RNN) architectures, this paper explores DNNs with the attention mechanism \cite{Transformer} at the encoder and decoder. Our main contributions are summarized as follows.
\begin{enumerate}
\item We present AttentionCode, a new class of DL-enab\-led feedback codes. Three underpinnings of  AttentionCode are AttentionNet, input restructuring, and adaptation to fading channels.
AttentionNet is adapted from the transformer architecture \cite{Transformer}, and leverages the attention mechanism to create temporal correlations at the transmitter for encoding, and exploits these temporal correlations at the receiver for decoding.
Input restructuring reorganizes the structure of the input feature matrix at the encoder (including both source messages and feedback) to enable AttentionNet better exploit the column-wise correlations to generate codewords.
We generalize AttentionCode to fading channels, wherein the receiver faces dramatically varying SNR due to fading.
\item DL for end-to-end wireless communication modeling is fundamentally different from classical machine learning tasks as the training and test datasets are generated by exactly the same distribution, and a test example may have be seen in the training phase.
To train a good model of AttentionCode, we put forth several training methodologies, including large-batch training, distributed learning, look-ahead optimizer, training-test SNR mismatch, and curriculum learning.
These training methods are shown to be empirically effective for AttentionCode and potentially beneficial to general DL applications in the area of wireless communications.

\item Numerical experiments verify that AttentionCode establishes a new state of the art among DL-based feedback codes.
On the one hand, this demonstrates the feasibility and superiority of pure-attention-based DNNs in the design of feedback codes, opening up a new prospect for general channel code design with the attention mechanism.
On the other hand, it validates the feasibility of utilizing feedback to achieve ultra-low BLER, demonstrating the potential of AttentionCode to provide ultra-reliable short-packet communications for future wireless communication systems. 
\end{enumerate}


{\it Notations} -- We use boldface lowercase letters to denote column vectors (e.g., $\bm{x}$, $\bm{y}$) and boldface uppercase letters to denote matrices (e.g., $\bm{S}$, $\bm{H}$). For a vector or matrix, $(\cdot)^\top$ denotes the transpose, $(\cdot)^*$ denotes the complex conjugate, and $(\cdot)^H$ denotes the conjugate transpose. $\|\cdot\|_0$ denotes the zero norm of a vector.
$\mathcal{R}$ and $\mathcal{C}$ stand for the sets of real and complex numbers, respectively. $(\cdot)^\mathfrak{r}$ and $(\cdot)^\mathfrak{i}$ stand for the real and imaginary components of a complex symbol or vector, respectively. The imaginary unit is represented by $j$. $\mathcal{N}$ and $\mathcal{CN}$ stand for the real and complex Gaussian distributions, respectively. The cardinality of a set $\mathcal{V}$ is denoted by $|\mathcal{V}|$.
$\text{round}(x)$ denotes the rounding operator that returns the nearest integer to $x$.
The modulo operation is defined as $\mathbb{M}_d[x]\triangleq x-d~\text{round}(x/d)$.

\section{Prior works}\label{sec:II}
Prior works on feedback-code designs often focus on the AWGN-channel case, wherein the feedforward and feedback channels are modeled as AWGN channels.
Consider the communications between two nodes A and B, where node A aims to transmit a bitstream $\bm{b}\in\{0,1\}^K$ to node B.
The feedforward and feedback transmissions can be represented by
\begin{eqnarray}
\label{eq:AWGN_fwd}
\bm{y}\hspace{-0.2cm}&=&\hspace{-0.2cm}\bm{x}+\bm{w}, \\
\label{eq:AWGN_fb}
\bm{y}^\prime\hspace{-0.2cm}&=&\hspace{-0.2cm}\bm{y}+\bm{w}^\prime\triangleq \bm{x}+\overline{\bm{w}},
\end{eqnarray}
where $\bm{x}\in\mathcal{R}^N$ consists of  encoded symbols at node A (i.e., a codeword);
$\bm{y}^\prime\in\mathcal{R}^N$ consists of the received feedback from node B;
$\bm{w}\in\mathcal{R}^N$ and $\bm{w}^\prime\in\mathcal{R}^N$ are real AWGN vectors, and their elements follow $\mathcal{N}(0,\widetilde{\sigma}^2_f)$ and $\mathcal{N}(0,\widetilde{\sigma}^2_b)$, respectively;
$\overline{\bm{w}}=\bm{w}+\bm{w}^\prime$ consists of the aggregated noise realizations observed at node A.
Nodes A and B are subjective to power constraints $P$ and $P'$, respectively.


\subsection{The SK and Modulo-SK schemes}
The seminal work of Schalkwijk and Kailath developed the SK scheme \cite{SK1,SK2}, for the AWGN-channel case with noiseless feedback, i.e., $\widetilde{\sigma}^2_b=0$.

The SK scheme breaks the transmission into two phases.
In the initialization phase, node A maps the bitstream $\bm{b}$ to a constellation point $\Theta$ of a normalized $2^K$-ary pulse-amplitude modulation (PAM) and transmits $\Theta$ to node B via the noisy feedforward channel. Node B computes an estimate of $\Theta$, denoted by $\hat{\Theta}_1$.
In successive interactions, node B feeds back its current estimate $\hat{\Theta}_n$ to node A via the noiseless feedback channel, from which node A knows the current estimation error of node B, i.e., $u_n=\hat{\Theta}_n-\Theta$. To decrease node B's estimation error, node A transmits a scaled version of $u_n$ to node B via the noisy feedforward channel.
Finally, after $N$ interactions, node B decodes the transmitted PAM constellation by the minimum distance decoding.
With the SK scheme, the authors proved that the decoding error of node B decays doubly exponentially fast in $N$ for any rate below capacity.

The SK scheme was designed for the noiseless feedback setting, wherein node A perfectly observes the estimation error of node B. It, however, fails when arbitrarily low noise is present in the feedback channel \cite{SK2,Kim2007ISIT}. In this context, the authors in \cite{ModuloSK} proposed a simple and efficient scheme, dubbed the modulo-SK scheme, to tackle noisy feedback channel.

The modulo-SK scheme made two major modifications to the interaction phase of the SK scheme:
1) Instead of $\hat{\Theta}_n$, node B feeds back $x'[n]=\mathbb{M}_d[\gamma_n\hat{\Theta}_n]$, i.e., a scaled version of $\hat{\Theta}_n$, modulo a fixed interval, to node A,
where $\gamma_n$ is a scaling factor and $d$ is set to $\sqrt{12P'}$ to ensure that node B's power constraint is satisfied.
2) Upon receiving $y'[n]=x'[n]+w'[n]$, node A estimates the estimation error of node B by $u_n=\mathbb{M}_d[y^\prime[n]-\gamma_n\Theta]$.
Compared with the SK scheme that completely fails when the feedback channel is noisy, the modulo-SK scheme works well when the feedback channel SNR is significantly larger than the feedforward channel SNR.
 

Due to the necessity of transmitting a $2^{K}$-ary PAM, the number of bits required by the SK and modulo-SK schemes to represent all statistics grows linearly with $K$. When $K$ is large, both schemes suffer from severe quantization errors caused by the finite-precision arithmetic and finite quantization levels of the electronic parts and components, e.g., power amplifier (PA) and field-programmable gate array (FPGA) chip.

When the information block length is $K=50$, for example, the first transmission of the SK and modulo-SK schemes requires at least 50 bits to represent the PAM modulation as well as all the other statistics involved in the arithmetic.
This, however, is challenging for state-of-the-art communication systems with finite-precision circuits and chips. Take the analog-to-digital converter (ADC) for instance. The most advanced ADC products (e.g., Texas Instruments ADS1263 and Analog Devices AD4134) support at most $32$-bit quantization.

\subsection{Deep learning-based schemes}
The pioneering work of DeepCode \cite{DeepCode} proposed a learning-based feedback code for AWGN channels, wherein the DNN encoder and decoder are chosen to be RNNs \cite{RNN} with the gated recurrent unit (GRU) \cite{GRU}.

DeepCode consists of two transmission phases. 
In the first phase, the source bitstream $\bm{b}$ is BPSK modulated and transmitted to node B in $K$ forward channel uses.
No feedback is utilized for encoding at node A.
The second phase, on the other hand, is a forward-coding phase with passive feedback.
In the $k$-th interaction of the second phase, $k=1,2,...,K$, node A generates two coded symbols\footnote{Unlike traditional feedforward channel codes, DL-based feedback codes generate discrete-time continuous amplitude coded symbols, which is known as discrete-time analog transmission \cite{DTAT}.} using the hidden states of GRU and four new symbols: the $k$-th bit $b_k$; the noise realization encountered by $b_k$ in the first phase; and the noise realizations encountered by the last two coded symbols of the second phase.
After $K$ interactions, node B receives in total $2K$ real symbols in the second phase.
Finally, given the $3K$ symbols received from the two phases, node B decodes the bitstream by a RNN decode with bi-directional GRUs.

In the final design of DeepCode, the authors introduced two techniques, dubbed {\it zero padding} and {\it power reallocation}, at node A to alleviate the unbalanced BER at different positions.
Specifically, zero padding pads an extra bit ``0'' at the end of sequence $\bm{b}$, and power reallocation reallocates power for encoded symbols at different positions by multiplying them with different weights while satisfying the power constraint.

Following the footprint of DeepCode, the authors in \cite{DEFC} proposed Deep Extended Feedback Codes (DEFC) to design feedback codes with better performance.
Three main improvements of DEFC over DeepCode are:
\begin{itemize}
\item {\it Higher-order modulation}. DEFC uses quadrature amplitude modulation (QAM) in the first phase, thereby achieving higher spectral efficiency than DeepCode.
\item {\it Long Short-Term Memory (LSTM)}. DEFC substitutes the GRUs at the encoder and decoder by LSTMs \cite{LSTM}, which can potentially provide longer-range dependencies.
\item {\it Feedback extension}. DEFC generates parity symbols at node A based on received feedback over a longer time window, thereby manually introducing longer-range dependencies between parity symbols.
\end{itemize}

The authors in \cite{DRFC} proposed Deep SNR-Robust Feedback Codes (DRFC)  that are robust to SNR mismatches between the training and test SNRs (instantaneous link SNR).
\begin{itemize}
\item The main idea of DRFC is to let the decoder be aware of the feedforward and feedback channel SNRs by feeding $\widetilde{\sigma}^2_f$ and $\widetilde{\sigma}^2_b$ into the decoder as additional information to aid training. As a result, the trained neural encoder and decoder are adaptive in a range of test SNRs.
\item The author extended the AWGN-channel case to a semi-fading case, where the feedforward channel is a fading channel while the feedback channel is an AWGN channel. Moreover, the fading feedback channel considered in \cite{DRFC} is limited to magnitude fading and the phase offsets are assumed to be perfectly compensated.
\end{itemize}

\section{System Model}\label{sec:III}
\begin{figure}[t]
  \centering
  \includegraphics[width=0.9\columnwidth]{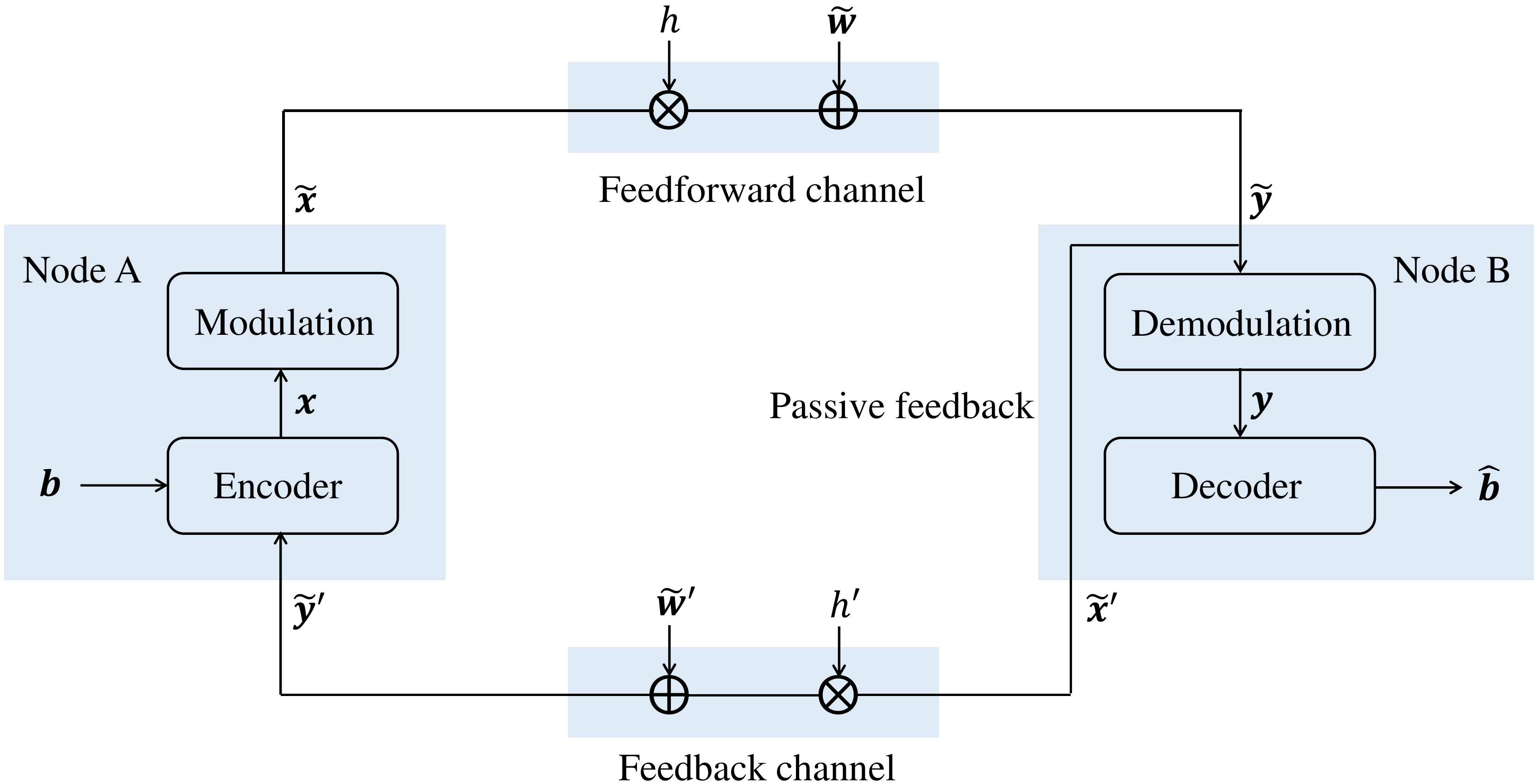}\\
  \caption{A pair of nodes A and B collaborate to communicate via the feedforward and feedback channels. One interaction between nodes A and B consists of one feedforward transmission from node A to B and one feedback transmission from node B to A.}
\label{fig:model}
\end{figure}

\begin{table}[t]
\caption{Main notations.}
\label{tab:notations}
\centering
\setlength{\tabcolsep}{3mm} 
\begin{tabular}{cc}
\toprule
Description               & Symbol \\ 
\midrule
Source bitstream        &  $\bm{b}\in\{0,1\}^K$ \\   
Codeword                &  $\bm{x}=\{x[n]:n=1,2,...,N\}$ \\
Number of interactions  &  $\widetilde{N}=\lceil N/2\rceil$    \\
Node A's transmitted symbols & $\widetilde{\bm{x}}=\{\widetilde{x}[\widetilde{n}]:\widetilde{n}=1,2,...,\widetilde{N}\}$\\
Node B's received symbols &   $\widetilde{\bm{y}}=\{\widetilde{y}[\widetilde{n}]:\widetilde{n}=1,2,...,\widetilde{N}\}$\\
Node B's received codeword &  $\bm{y}=\{y[n]:n=1,2,...,N\}$\\
Node B's feedback symbols           &  $\widetilde{\bm{x}}^\prime=\{\widetilde{x}^\prime[\widetilde{n}]:\widetilde{n}=1,2,...,\widetilde{N}-1\}$ \\
Node A's received feedback           &  $\widetilde{\bm{y}}^\prime=\{\widetilde{y}^\prime[\widetilde{n}]:\widetilde{n}=1,2,...,\widetilde{N}-1\}$                                                  \\ 
\bottomrule
\end{tabular}
\end{table}

We consider a point-to-point communication system with two nodes A and B, as shown in Fig. \ref{fig:model}. The two nodes are connected via a feedforward channel and a feedback channel, both of which are modeled as Rayleigh fading channels.
Specifically, suppose node A transmits a vector of complex symbols $\widetilde{\bm{x}}\in\mathcal{C}^{\widetilde{N}}$, then the received complex symbols at node B are given by
\begin{equation}\label{eq:fwd_complex}
\widetilde{\bm{y}} = h\widetilde{\bm{x}} +\widetilde{\bm{w}},
\end{equation}
where i) the Rayleigh fading channel coefficient $h\sim\mathcal{CN}\allowbreak(0,\allowbreak 2\widetilde{\rho}^2_f)$; and
ii) $\widetilde{\bm{w}}$ is an additive white Gaussian noise (AWGN) vector, the elements of which are sampled from a complex Gaussian distribution $\mathcal{CN}\allowbreak(0,\allowbreak 2\widetilde{\sigma}^2_f)$ in an independent and identically distributed (i.i.d.) manner.
We consider slow fading, and hence, $h$ remains constant over the transmission of $\widetilde{\bm{x}}$.
Likewise, if we denote by $\widetilde{\bm{x}}^\prime\in\mathcal{C}^{\widetilde{N}}$ the complex symbols transmitted by node B, the received symbols at node A are given by
\begin{equation}\label{eq:fb_complex}
    \widetilde{\bm{y}}^\prime = h^\prime\widetilde{\bm{x}}^\prime +\widetilde{\bm{w}}^\prime,
\end{equation}
where $h^\prime\sim\mathcal{CN}(0,2\widetilde{\rho}^2_b)$, and the elements of $\widetilde{\bm{w}}^\prime$ follow $\mathcal{CN}(0,2\widetilde{\sigma}^2_b)$.
We assume that the channel state information (CSI) $\{h, h^\prime\}$ is known to both nodes A and B.
Moreover, nodes A and B have average power constraints $2P$ and $2P^\prime$, respectively. The power constraints will be defined precisely later on.

Time is divided into slots. One interaction between nodes A and B takes one time slot and consists of one feedforward transmission from node A to B and one feedback transmission from node B to A. To ease reading, we summarize the main notations in Table~\ref{tab:notations}.

In this system, the communication goal is to deliver a vector of $K$ bits $\bm{b}\in\{0,1\}^K$ reliably from node A to B. 
To this end, nodes A and B communicate over $\widetilde{N}$ interactions.

In the $\widetilde{n}$-th interaction, $\widetilde{n}=1,2,...,\widetilde{N}$,
\begin{enumerate}[i)]
\item \textbf{Channel coding}: Node A generates 2 coded symbols, $x[2\widetilde{n}-1]$, $x[2\widetilde{n}]\in\mathcal{R}$, based on the source bitstream $\bm{b}$ and all the complex feedback symbols received so far, i.e., $\{\widetilde{y}^\prime[\widetilde{n}]:\widetilde{n}=1,2,...,\widetilde{n}-1\}$.
\item \textbf{Forward transmission}: Node A modulates $x[2\widetilde{n}-1]$ and $x[2\widetilde{n}]$ onto a carrier frequency as the in-phase and quadrature components, respectively, yielding $\widetilde{x}[\widetilde{n}]=x[2\widetilde{n}-1]+jx[2\widetilde{n}]$.
Then, node A transmits $\widetilde{x}[\widetilde{n}]$ to node B through the feedforward channel and node B receives $\widetilde{y}[\widetilde{n}]$, following \eqref{eq:fwd_complex}.
\item \textbf{Feedback}: Given the received symbol, node B feeds back a symbol $\widetilde{x}^\prime[\widetilde{n}]$ to node A. There are two feedback mechanisms in the literature: passive feedback \cite{Shannon,Kim2007ISIT,DeepCode} and active feedback \cite{SK1,ModuloSK}. This paper considers passive feedback, in which case the feedback symbol is the raw received symbol, i.e., $\widetilde{x}^\prime[\widetilde{n}]=\widetilde{y}[\widetilde{n}]$.
After passing through the feedback channel, node A receives $\widetilde{y}^\prime[\widetilde{n}]$, as defined in \eqref{eq:fb_complex}. We note that there is no feedback for the final $\widetilde{N}$-th forward transmission.
\end{enumerate}

After $\widetilde{N}$ interactions, node B collects $\widetilde{N}$ complex symbols $\widetilde{\bm{y}}$ and forms a vector of $N$ real symbols $\bm{y}$, based on which node B aims to decode the transmitted bitstream.

Denote by $\widehat{\bm{b}}$ the decoded bitstream, we are interested in the block error rate (BLER) $P_B=\Pr(\bm{b}\neq\widehat{\bm{b}})$ as a function of the signal-to-noise ratio (SNR) of feedforward and feedback channels.
Let\footnote{Here, we define $\eta^\prime=\frac{P}{\widetilde{\sigma}^2_b}$ as opposed to $\frac{P^\prime}{\widetilde{\sigma}^2_b}$ to be consistent with prior works \cite{DeepCode,DEFC,DRFC}.}
\begin{eqnarray}\label{eq:SNRs}
\eta\triangleq \frac{P}{\widetilde{\sigma}^2_f},~~~~~~
\eta^\prime\triangleq \frac{P}{\widetilde{\sigma}^2_b}.
\end{eqnarray}
The SNRs of the feedforward and feedback channels are defined, respectively, as
\begin{equation}\label{eq:SNRRs}
\eta_r\triangleq\frac{P\lvert h \rvert^2}{\widetilde{\sigma}^2_f}= \lvert h \rvert^2\eta, ~~
\eta^\prime_r\triangleq\frac{P\lvert h \rvert^2\lvert h^\prime \rvert^2}{\widetilde{\sigma}^2_b}=\lvert h \rvert^2\lvert h^\prime \rvert^2\eta^\prime.
\end{equation}
Since $\lvert h \rvert^2\sim \text{Exp}(\frac{1}{2\widetilde{\rho}^2_f})$ and $\lvert h^\prime \rvert^2\sim \text{Exp}(\frac{1}{2\widetilde{\rho}^2_b})$, the average SNRs are given by $\mathbb{E}[\eta_r]= 2\widetilde{\rho}^2_f\eta$ and $\mathbb{E}[\eta^\prime_r]= 4\widetilde{\rho}^2_f\widetilde{\rho}^2_f\eta^\prime$.

Over the course of $\widetilde{N}$ interactions, node A transmits a codeword of $N=2\widetilde{N}$ real symbols  $\bm{x}\in\mathcal{R}^{N}$; hence,
the code rate is $R=K/N$. 
We impose the following constraint on $\bm{x}$ to meet the  power constraint:
\begin{equation}\label{eq:fwd_power}
\frac{1}{N}\mathbb{E}\left(\sum_{n=1}^{N}x[n]^2\right)\leq P.
\end{equation}
Provided that \eqref{eq:fwd_power} is satisfied, it is easy to verify that the symbols fed back by node B also satisfy a power constraint of $4\widetilde{\rho}^2_fP+2\widetilde{\sigma}^2_f$ with the passive feedback scheme. 
Therefore, we assume the average power constraint of node B is $2P^\prime=4\widetilde{\rho}^2_fP+2\widetilde{\sigma}^2_f$.

\begin{rem}
If $2P^\prime\neq 4\widetilde{\rho}^2_fP+2\widetilde{\sigma}^2_f$, node B can simply scale the symbols to be transmitted by a factor of $\alpha^\prime=\sqrt{2P^\prime/(4\widetilde{\rho}^2_fP+2\widetilde{\sigma}^2_f)}$ to satisfy the power constraint. Correspondingly, node A scales the received symbols by $1/\alpha^\prime$, yielding a received signal similar to the case where $2P^\prime=4\widetilde{\rho}^2_fP+2\widetilde{\sigma}^2_f$ with the only difference being the noise power.
\end{rem}

\begin{rem}
When the noise power of the feedback channel is $2\widetilde{\sigma}^2_b=0$, there is no difference between passive and active feedback, because any active processing at node B can be equivalently moved to node A thanks to the noiseless feedback channel.
An application of such a noiseless feedback system is the ground-satellite communication: the ground-to-satellite direction has a much larger transmission power than the reverse link, and hence, can be taken as noiseless \cite{SK1}.
\end{rem}

To conclude this section, we comment on some practical concerns of the interaction-based communication paradigm. The feedback from the receiver incurs additional overhead on latency, energy, and computation. Here, we focus on the latency and energy costs, while the computational complexity of AttentionCode will be discussed later in Remark \ref{rem:comp}. 

\textbf{Latency} --
In commercial cellular systems, e.g., 4G and 5G, communications operate in time slots. A time slot consists of several orthogonal frequency-division multiplexing (OFDM) symbols, where some OFDM symbols are assigned to uplink and others are assigned to downlink. 
Suppose the transmission from a user to a base station lasts for $\widetilde{N}$ time slots. With the conventional design, the user only utilizes the uplink resource for data transmissions and the delay is $\widetilde{N}$ time slots. With AttentionCode, in contrast, the base station can utilize the downlink resource (which is often richer than the uplink) and feedback on some information to improve the uplink transmission. The delay is still $\widetilde{N}$ time slots.
The interaction-based communication paradigm naturally fits into existing communication systems and does not incur additional latency.

\textbf{Energy} --
Feedback incurs additional energy consumption at the receiver. In communication systems, the energy sources of the transmitter and receiver are different. In uplink transmission, for example, the mobile user’s energy comes from the limited battery. The base station’s energy, on the other hand, is considered to be infinite and the additional energy consumption of feedback is not a problem. This suggests that AttentionCode is suitable for scenarios where the receiver is not energy-constrained.

\section{AttentionCode: Architectural Innovations}\label{sec:IV}
\begin{figure}[t]
    \centering
    \includegraphics[width=0.9\columnwidth]{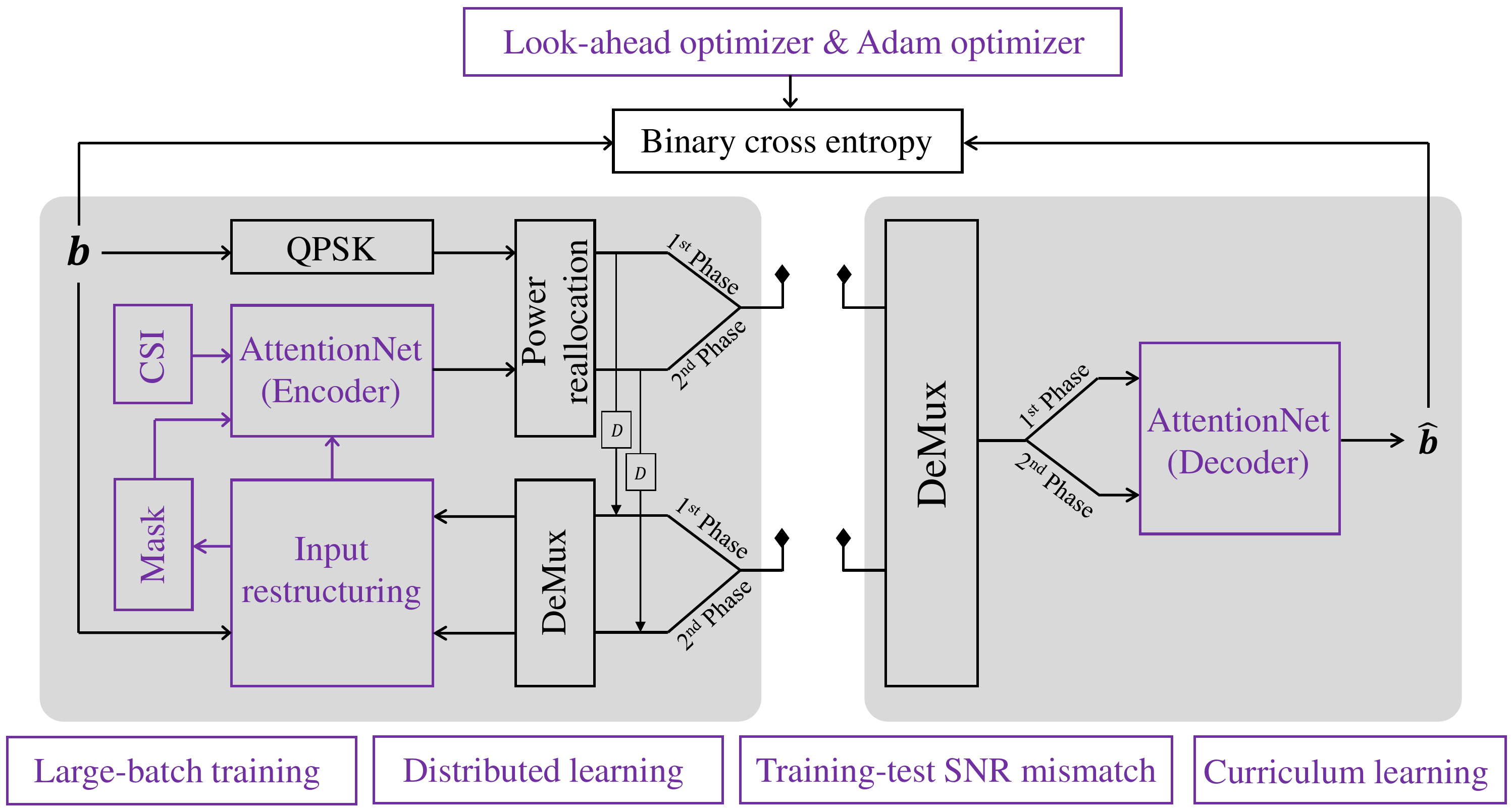}
    \caption{The architecture of AttentionCode with two transmission phases. The innovations over existing DL-based feedback-code designs are highlighted in blue.}
    \label{fig:innovations}
\end{figure}

This paper presents AttentionCode, a new class of feedback codes based on DL architecture.
Existing DL-based feedback-code designs rely exclusively on the recurrent encoder and decoder structures with GRU or LSTM as the recurrent units. AttentionCode, in contrast, explores a new network architecture, which is based purely on the attention mechanism \cite{Transformer}, to design feedback codes. To start with, we outline the architecture of AttentionCode in Fig.~\ref{fig:innovations} and highlight our innovations over existing DL-based designs. 

In the big picture, AttentionCode adopts the same two-phase transmission paradigm with passive feedback as DeepCode.
As shown in Fig.~\ref{fig:innovations}, in the first phase, node A transmits quadrature phase shift keying (QPSK)-modulated symbols to node B in an uncoded fashion.
Following \eqref{eq:AWGN_fwd} and \eqref{eq:AWGN_fb}, let us denote the received passive feedback at node A by
$y'_{0,k} \triangleq  x_{0,k}+\overline{w}_{0,k}$, $k=1,2,...,K$,
where `0' in the subscript denotes the first phase, $x_{0,k}$ is the QPSK symbol, and $\overline{w}_{0,k}$ is the aggregated noise realization encountered by $x_{0,k}$.

In the second phase, the encoder sequentially generates $2K$ real coded symbols  and transmits them in $\widetilde{N}=K$ interactions.
In the $k$-th interaction, two real coded symbols $x_{1,k}$ and $x_{2,k}$ are generated, and we denote the feedback symbols received at node A by
$y'_{1,k} \triangleq  x_{1,k}+\overline{w}_{1,k}$ and
$y'_{2,k} \triangleq x_{2,k}+\overline{w}_{2,k}$,
where $\overline{w}_{1,k}$ and $\overline{w}_{2,k}$ are the aggregated noise realizations encountered by the first and second coded symbols, respectively. Since $x_{1,k}$ and $x_{2,k}$ are known to node A, $\overline{w}_{1,k}$ and $\overline{w}_{2,k}$ can be obtained for the encoding of $x_{1,k+1}$ and $x_{2,k+1}$ in the next interaction.

After the two transmission phases, node B feeds all three received sequences $\{y_{0,k},y_{1,k},y_{2,k}:\forall k \}$ into the decoder. The encoder and decoder networks are jointly trained to minimize the binary cross-entropy loss between the transmitted bit stream $\bm{b}$ and the estimated probabilities:
\begin{equation}\label{eq:loss}
\mathcal{L}\!=\!\frac{1}{K}\sum_{k=1}^{K}\!\Big[-b_k\log{\Pr(b_k\!=\!1)}-(1-b_k)\log\big(1-\Pr(b_k\!=\!1)\big)\Big].
\end{equation}

Our main architectural innovations are threefold:
\begin{itemize}
\item {AttentionNet}: We substitute RNNs used in prior works by a class of pure-attention-based DNNs, dubbed AttentionNet, as the encoder and decoder.
\item {Input restructuring}: We restructure the information available for encoding in each interaction so that the noise realizations encountered by a single bit are aligned in one column of the input matrix. To enable better learning performance, we leverage ``masks'' to constrain the input columns that a coded symbol attends to.
\item {Adaptation to fading channels}: We extend the DL architecture to fading channels.
The inputs to the encoder and decoder are modified and the channel state information (CSI) is incorporated to adapt to fading.
\end{itemize}

This section focuses on these three architectural innovations. The training methods presented in Fig.~\ref{fig:innovations} will be treated in more detail in Section~\ref{sec:V}.

\subsection{AttentionNet}\label{sec:IIIA}
RNNs, with various embedded gate units (e.g., GRU and LSTM), have been successfully used for decades for inference on temporally correlated data.
In 2017, the authors of \cite{Transformer} introduced a pure-attention-based DNN architecture, dubbed ``transformer'', to replace the prevalent RNN models for machine translation tasks. Since then, pure-attention-based DNN models are getting more and more popular in the DL community, while also finding applications in wireless communications \cite{transformer1}.

Transformer is a sequence-to-sequence DNN model. It is composed of several encoding  and decoding blocks \cite{Transformer}: the encoding blocks take the input sequence and map it into a higher-dimensional feature vector; the decoding blocks, on the other hand, take the feature vector as the input and decode it to the target sequence.
In AttentionCode, we utilize a simplified version of the transformer, dubbed {\it AttentionNet}. In particular, AttentionNet is modified from the encoding blocks of the transformer and does away with the decoding blocks. In the literature, a similar model is known as bidirectional encoder representations from transformers (BERT) \cite{BERT}. However, BERT often refers to the pre-trained language models using unsupervised learning. To avoid ambiguity, here we refer to the pure-attention-based DNNs with the encoding blocks of transformer as AttentionNet.

\begin{figure}
    \centering
    \includegraphics[width=0.9\columnwidth]{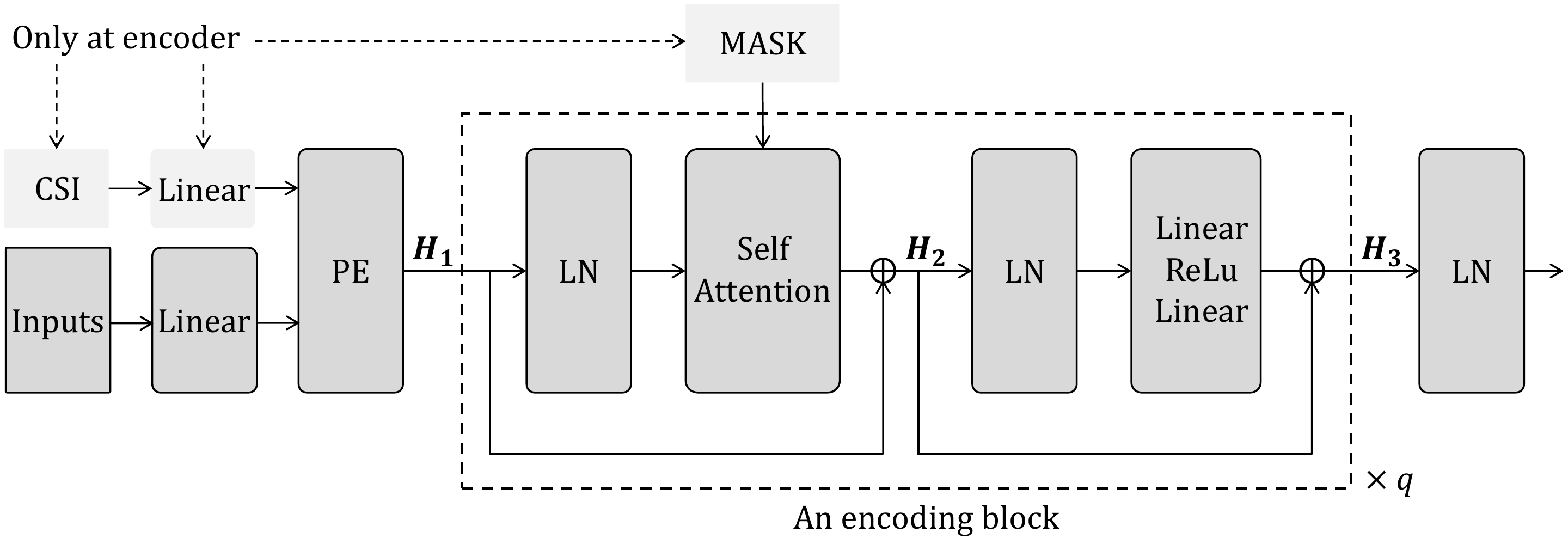}
    \caption{The detailed structure of AttentionNet, where ``PE'' stands for positional encoding and ``LN'' stands for layer normalization. The ``CSI'', ``Linear'' and ``MASK'' modules (colored in light gray) exist only when AttentionNet is used as encoder at node A.}
    \label{fig:AttentionNet}
\end{figure}

The detailed structure of AttentionNet is presented in Fig.~\ref{fig:AttentionNet}.
In this subsection, we will consider the AWGN-channel case, and hence, ignore the ``CSI'' and ``Linear'' modules on the left-hand side (LHS) of Fig.~\ref{fig:AttentionNet} (colored in light grey); they will be included in the more general fading-channel case in Section~\ref{sec:IIIC}.
To illustrate the operations inside AttentionNet, we focus on AttentionNet as the encoder and assume that the input to AttentionNet is the same as DeepCode. The input matrix will be restructured later in Section~\ref{sec:IIIB}.

Let us start from the ``Inputs'' module of Fig.~\ref{fig:AttentionNet}. Denote by $\bm{S}(k)$ the input matrix to the encoder in the $k$-th interaction of the second phase. We have
\begin{eqnarray}\label{eq:input1}
\renewcommand*{\arraystretch}{0.8}
\bm{S}(k)=\begin{bmatrix}
b_1 & b_2 & b_3 & ... & b_k \\
\overline{w}_{0,1} & \overline{w}_{0,2} & \overline{w}_{0,3} & ... & \overline{w}_{0,k}\\ 
0 & \overline{w}_{1,1} & \tilde{w}_{1,2} & ... & \overline{w}_{1,k-1}\\ 
0 & \overline{w}_{2,1} & \overline{w}_{2,2} &  ... & \overline{w}_{2,k-1}
\end{bmatrix}.
\end{eqnarray}
In the $k$-th interaction, $k=1,2,...,K+1$, the dimension of  $\bm{S}(k)$ is ${d_S\times k}$, where $d_S=4$ when AttentionNet is used as the encoder and $d_S=3$ when AttentionNet is used as the decoder (because the input to the decoder consists of three sequences of received symbols: $\{y_{0,k},y_{1,k},y_{2,k}:\forall k\}$). 

The input matrix $\bm{S}$ will first be passed through a linear layer for feature extraction and a positional encoding (PE) layer, yielding
\begin{eqnarray}
\bm{H_1} = \bm{L_1 S + W_{\text{pos}}},
\end{eqnarray}
where $\bm{L_1}\in \mathcal{R}^{d_m\times d_S}$ represents the weight matrix of the linear layer; $\bm{W}_{\text{pos}}\in\mathcal{R}^{d_m\times k}$ is a constant matrix, representing the positional information embedded in the input matrix (we use the same $\bm{W_{\text{pos}}}$ as \cite{Transformer});\footnote{The self-attention model treats each column of $\bm{L_1 S}$ as independent of the other and cannot recognize the sequential order of the columns. As a result, positional encoding explicitly adds the positional information to the input such that the knowledge about the order of columns is maintained in the input.} and $d_m$ is a variate that determines the model size.

Then, $\bm{H_1}$ will go through $q$ encoding blocks.
For each encoding block, we adopt a pre-normaliza\-tion framework proposed in \cite{PerNorm}. That is, input data are first layer-normalized \cite{LN} before they are fed into the self-attention module or the linear module, as shown in Fig.~\ref{fig:AttentionNet}.
Let us abstract a single encoding block as a function $g$, then the final output of the encoding layers is
$\bm{H_3} = g^{q}(\bm{H_1})$.
Since the encoding block is the most essential part of the AttentionNet, we will focus on how the encoding block works in the following. 

To ease exposition, we assume $q=1$ and explain how to get $\bm{H_3}$ from $\bm{H_1}$.
The processing in one encoding block can be written as
\begin{eqnarray}
&& \bm{H_2} = \text{SelfAttention}(\text{LN}(\bm{H_1})) + \bm{H_1}, \\
&& \bm{H_3} = \bm{L_3}\text{ReLu}(\bm{L_2}(\text{LN}(\bm{H_2}))) + \bm{H_2},
\end{eqnarray}
where ``$\text{LN}$'' stands for the layer normalization module; $\text{SelfAttention}$  stands for the self-attention module, which will be detailed later;  $\bm{L_2}\in\mathcal{R}^{4d_m\times d_m}$ and $\bm{L_3}\in\mathcal{R}^{d_m\times 4d_m}$ are the weight matrices of the linear layers, respectively. Note that $\text{LN}$ and $\text{SelfAttention}$ modules do not change the shape and size of their inputs.

The output $\bm{H_3}\in\mathcal{R}^{d_m\times k}$ will be layer normalized and used to generate the encoded symbols (at the encoder) or decode the bitstream (at the decoder). This can be realized simply by a linear transformation, i.e., a fully-connected layer with weights of size $\mathcal{R}^{2\times d_m}$.

The essential component of the encoding block is the self-attention module. Let the input and output of the self-attention module be matrices $\bm{J}$ and $\bm{J^\prime}$, respectively.
Then, $\bm{J^\prime}$ is the features extracted from $\bm{J}$ and they have the same size.
In particular, the $i$-th column of $\bm{J^\prime}$ is a linear combination of $\bm{J}$'s columns, and the combination weights is determined by the inter-correlations between the $i$-th column of $\bm{J}$ and the other columns of $\bm{J}$.

More rigorously, let $\bm{J}=\text{LN}(\bm{H_1})\in\mathcal{R}^{d_m\times k}$. We first generate a query matrix $\bm{Q}$, a key matrix $\bm{K}$, and a value matrix $\bm{V}$ \cite{Transformer} using three weight matrices $\bm{W^q}\in\mathcal{R}^{d_m\times d_m}$, $\bm{W^k}\in\mathcal{R}^{d_m\times d_m}$,  $\bm{W^v}\in\mathcal{R}^{d_m\times d_m}$,  respectively, where $\bm{Q=W^q J}$, $\bm{K=W^k J}$ and $\bm{V=W^v J}$.
A correlation matrix capturing the inter-column correlations of $\bm{J}$ can be computed by
$\bm{C=Q^\top K}$.

For a given mask matrix $\bm{M}\in\mathcal{R}^{k\times k}$, the masked corre\-lation matrix is given by
$\bm{C^\prime = C \odot M}$,
where $\odot$ is the element-wise product.
A commonly used mask in transformer is called the causal mask given by
\begin{equation}\label{eq:M}
\renewcommand*{\arraystretch}{0.6}
\bm{M}=
\begin{bmatrix}
 1 & -\infty & -\infty & ... & -\infty \\ 
 1 & 1 & -\infty & ... & -\infty \\ 
 1 & 1 & 1 & ... & -\infty \\ 
 1 & 1 & 1 & ... & 1
\end{bmatrix}_{(K+1)\times(K+1)}.
\end{equation}

The final output of the self-attention block is obtained as
\begin{equation}\label{eq:Jprime}
\bm{J^\prime} =\bm{V}\text{softmax}(\bm{C}').
\end{equation}
Note that after passing through the $\text{softmax}$ operation in \eqref{eq:Jprime}, $\bm{M}$ becomes a left triangular matrix.

\begin{figure}
    \centering
    \includegraphics[width=0.9\columnwidth]{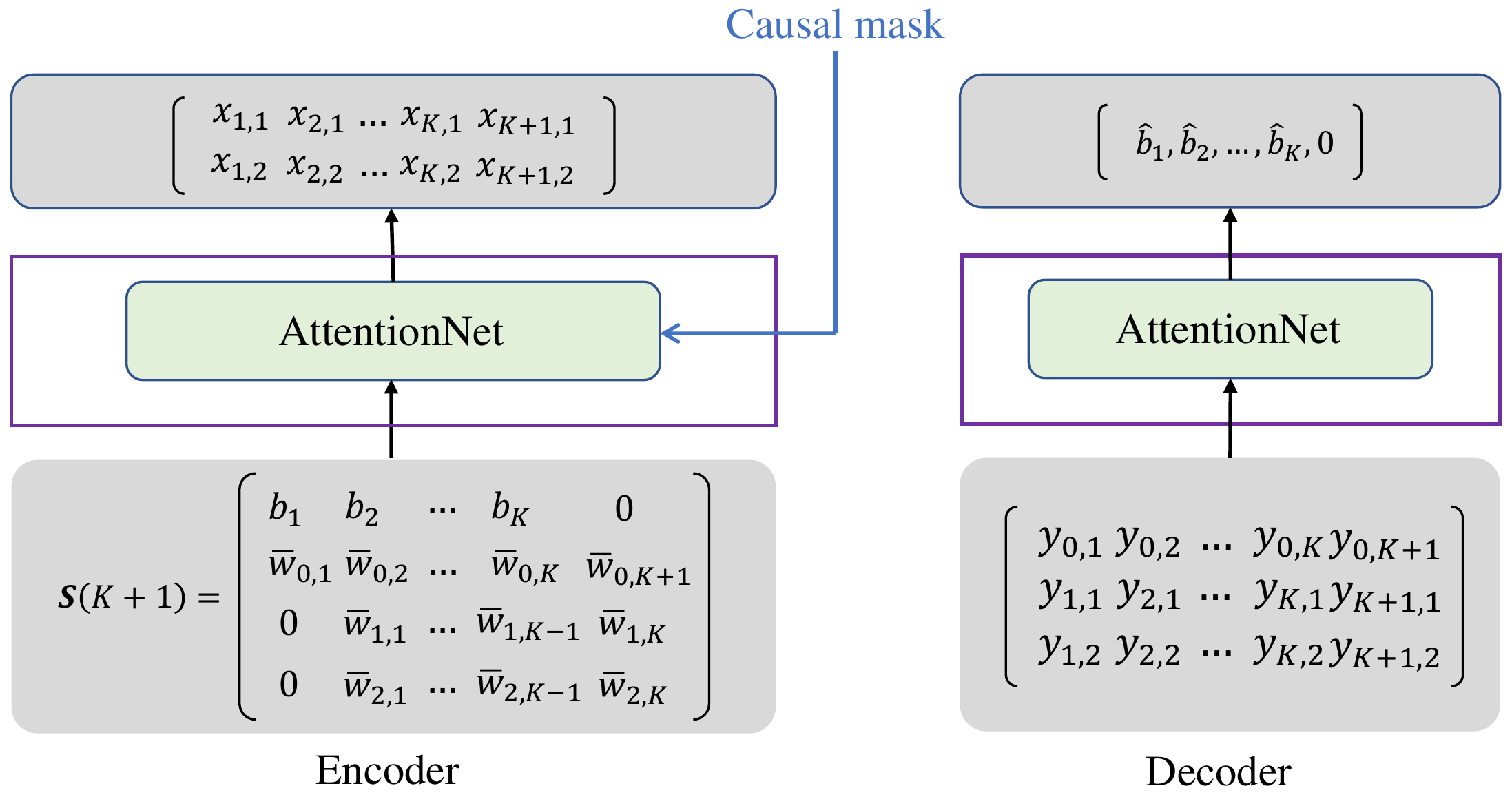}
    \caption{Substituting AttentionNet for RNN in DeepCode.}
    \label{fig:design1_simplified}
\end{figure}

The causal mask enables the parallel computation of coded symbols in the training phase. As shown in Fig.~\ref{fig:design1_simplified}, we can feed all the information (i.e., the matrix $\bm{S}(K+1)$) simultaneously into the AttentionNet so that all the coded symbols can be computed in a parallel fashion. In this process, a causal mask is essential since it guarantees that a coded column $[x_{k,1},x_{k,2}]^\top$ can only access the noise realizations that have already been observed, i.e., $\{\overline{w}_{0,k^\prime},\overline{w}_{1,k^\prime-1},\overline{w}_{2,k^\prime-1}:k^\prime=1,2,...,k\}$, but not those that will be observed in the future. 
In other words, albeit the real interactions among nodes A and B proceed sequentially, the training of Fig.~\ref{fig:design1_simplified} proceeds in parallel with the use of a ``causal mask'' to prevent future-information leakage, which makes training more efficient. This mechanism is similar to the uni-directional RNN in DeepCode.

\begin{rem}
With the simple substitution in Fig.~\ref{fig:design1_simplified}, AttentionCode achieves slightly better performance than DeepCode, but cannot beat state-of-the-art DL-based feedback codes such as DEFC and DRFC \cite{DEFC,DRFC}.
In the next subsection, we shall explore the idea of input restructuring to boost the performance of AttentionCode.
In particular, we will use masks in a different way to causal mask.
\end{rem}

\begin{rem}
Compared with different variations of transformer, AttentionNet does away with two widely-used techniques: one is the muti-head attention mechanism  and the other is dropout, because we  empirically found that they hinder AttentionCode from achieving low BLER.
The reason behind these empirical observations may be the fundamental difference between the channel coding problem and classical DL tasks that DL for channel coding does not always pursue the generalizability of DNNs, which will be explained in more detail in Section \ref{sec:V}.
\end{rem}

\begin{rem}
It is worth noting that AttentionNet serves two purposes:
i) as the encoder at node A to {\it create} temporal correlations among the encoded symbols;
ii) as the decoder at node B to {\it exploit} the temporal correlations among coded symbols, and hence, combat channel impairments such as noise and fading.
\end{rem}

\subsection{Input restructuring}\label{sec:IIIB}
As shown in Fig.~\ref{fig:design1_simplified}, if we replace the RNN by AttentionNet, the information used to generate $x_{k,1}$ and $x_{k,2}$ is the first $k$ columns of the input, i.e., the matrix $\bm{S}(k)$ given in \eqref{eq:input1}.
In particular, due to the delayed feedback, the last two rows of $\bm{S}(k)$ are delayed by one column compared with the first two rows.

In AttentionCode, we propose to reorganize $\bm{S}(k)$ so that the noise realizations encountered by a single bit in two transmission phases are aligned in one column of $\bm{S}(k)$. After restructuring, the new input matrix is given by
\begin{eqnarray}\label{eq:input2}
\renewcommand*{\arraystretch}{0.8}
\widetilde{\bm{S}}(k)=\begin{bmatrix}
b_1 & b_2 & b_3 & ... & b_{k-1} & b_k \\ 
\overline{w}_{0,1} & \overline{w}_{0,2} & \overline{w}_{0,3} & ... & \overline{w}_{0,k-1} & \overline{w}_{0,k}\\ 
\overline{w}_{1,1} & \overline{w}_{1,2}  & \overline{w}_{1,3} & ... & \overline{w}_{1,k-1} & 0 \\ 
\overline{w}_{2,1} & \overline{w}_{2,2} & \overline{w}_{2,3} & ... & \overline{w}_{2,k-1} & 0
\end{bmatrix}.
\end{eqnarray}

This scheme, dubbed ``input restructuring'', turns out to be very effective. 
One possible explanation is that the restructured input matrix $\widetilde{\bm{S}}(k)$ matches the nature of self-attention that focuses on the column-wise correlations among the inputs. 

Although input restructuring is a small modification to the input, it brings about a fundamental change in the system training from parallel to sequential. Specifically, without input restructuring, the difference between $\bm{S}(k)$ and $\bm{S}(k+1)$ is only the last column of $\bm{S}(k+1)$. Therefore, all $K+1$ columns can be fed into AttentionNet simultaneously, provided that a causal mask is used to prevent information leakage. In so doing, the overall training proceeds in parallel and all the coded symbols can be computed simultaneously. 

\begin{figure}[t]
    \centering
    \includegraphics[width=0.95\columnwidth]{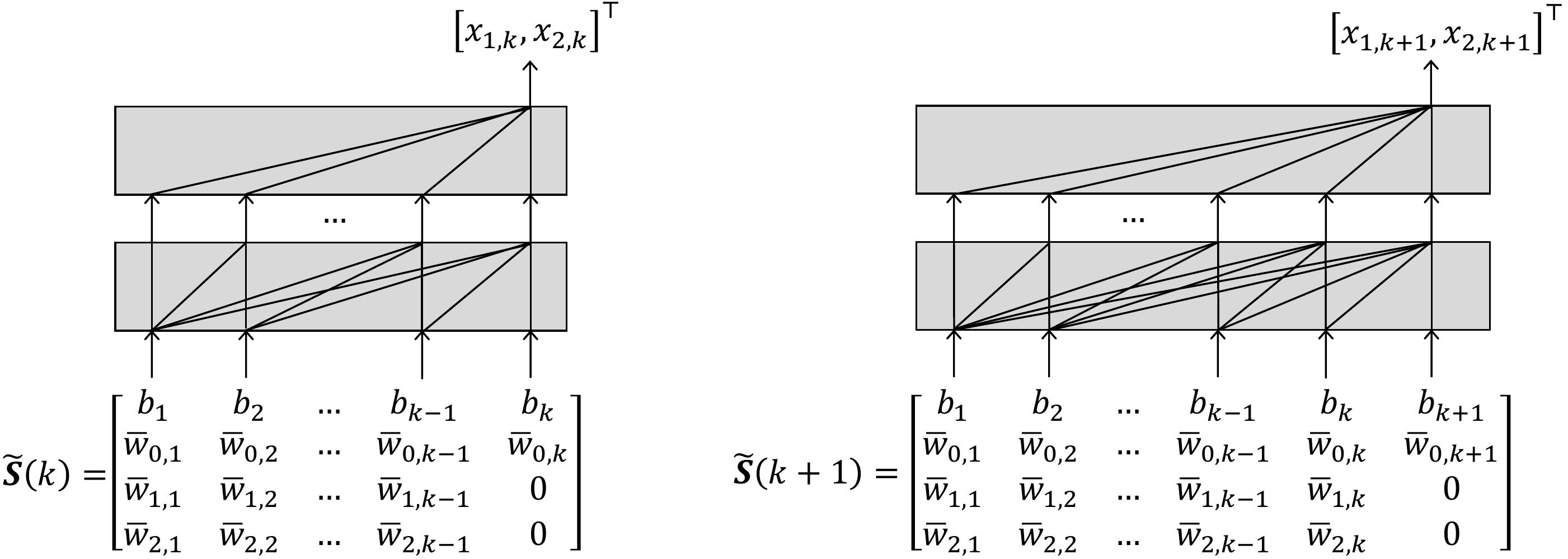}
    \caption{AttentionCode: an illustration of input restructuring with masks. The grey blocks stand for the self-attention module, wherein the arrows stand for the unmasked correlations to be computed.}
    \label{fig:inputRestructure}
\end{figure}

With input restructuring, on the other hand, $\widetilde{\bm{S}}(k+1)$ differs from $\widetilde{\bm{S}}(k)$ in last two columns. More importantly, unlike \eqref{eq:input1}, the generation of $\widetilde{\bm{S}}(k)$ in \eqref{eq:input2} is no longer recurrent -- the generation of $\widetilde{\bm{S}}(k+1)$ does not implicitly generate $\widetilde{\bm{S}}(k^\prime)$, $\forall k^\prime\leq k$.
As a consequence, AttentionCode can only be trained sequentially: in each interaction, we feed $\widetilde{\bm{S}}(k)$ into the AttentionNet and take the two output symbols of the last column as $[x_{1,k},x_{2,k}]^\top$.\footnote{Inside GPU, a concatenated computation graph is created for consecutive interactions.}
An illustration is given in Fig.~\ref{fig:inputRestructure}.

Another ingredient we use in conjunction with input restructuring is the mask. As shown in Fig.~\ref{fig:inputRestructure}, the self-attention layers are partially connected, subjected to a $k\times k$ dimensional mask
\begin{equation}
\renewcommand*{\arraystretch}{0.6}
\bm{M}_k=
\begin{bmatrix}
 1 & -\infty & -\infty & ... & -\infty \\ 
 1 & 1 & -\infty & ... & -\infty \\ 
 1 & 1 & 1 & ... & -\infty \\ 
 1 & 1 & 1 & ... & 1
\end{bmatrix}_{k\times k}.
\end{equation}

At first sight, $\bm{M}_k$ is a mask generated by the causal constraint since it is in the same form as \eqref{eq:M}. However, note that AttentionCode is trained sequentially and there is no causal constraint -- $\widetilde{\bm{S}}(k)$ is fully known to node A when generating $[x_{1,k},x_{2,k}]^\top$.
In other words, it is legitimate to remove the mask and use fully-connected self-attention layers in Fig.~\ref{fig:inputRestructure}. Nevertheless, we empirically found that masked self-attention layers lead to better feedback codes and we conjecture that this is thanks to the simplified correlations among the columns of $\widetilde{\bm{S}}(k)$ to generate $[x_{1,k},x_{2,k}]^\top$. In the final design of AttentionCode, we use masked self-attention at the encoder and unmasked self-attention at the decoder.

\begin{rem}
When the number of encoding blocks (which is also the number of self-attention modules) $q=1$, there is no difference between masked and unmasked AttentionCode since the generated  $[x_{1,i},x_{2,i}]^\top$ are the same (see Fig.~\ref{fig:inputRestructure} for a visual explanation).
\end{rem}

\subsection{Adaptation to fading channels}\label{sec:IIIC}
So far we have focused on the AWGN-channel to explain the principles of AttentionNet and input restructuring. In this subsection, we extend AttentionCode to fading channels.

In fading channels, the received signals at nodes A and B, i.e., \eqref{eq:AWGN_fwd} and \eqref{eq:AWGN_fb}, can be written as
\begin{eqnarray}
\label{eq:AWGN_fwd_equi}
\frac{1}{h}\widetilde{\bm{y}} \hspace{-0.2cm}&=&\hspace{-0.2cm}
\widetilde{\bm{x}} + \frac{1}{h}\widetilde{\bm{w}}
\triangleq \widetilde{\bm{x}} + \bm{w},\\
\label{eq:AWGN_fb_equi}
\frac{1}{h^\prime h}\widetilde{\bm{y}}^\prime \hspace{-0.2cm}&=&\hspace{-0.2cm}
\widetilde{\bm{x}} +\frac{1}{h}\widetilde{\bm{w}} +\frac{1}{h^\prime h}\widetilde{\bm{w}}^\prime
\triangleq \widetilde{\bm{x}} + \bm{w} + \bm{w}^\prime.
\end{eqnarray}
Thus, compared with AWGN channels, fading channels vary the feedforward and feedback noise powers from codeword to codeword. Specifically, denote the power of $\bm{w}$ and $\bm{w}^\prime$ by ${\sigma}^2_f$ and ${\sigma}^2_b$, respectively. We have
\begin{equation}
    {\sigma}^2_f=\frac{2\widetilde{\sigma}^2_f}{\lvert h \rvert^2}, ~~~~~~
    {\sigma}^2_b=\frac{2\widetilde{\sigma}^2_b}{\lvert h \rvert^2\lvert h^\prime \rvert^2},
\end{equation}
where $\lvert h \rvert^2\sim \text{Exp}(\frac{1}{2\widetilde{\rho}^2_f})$ and $\lvert h^\prime \rvert^2\sim \text{Exp}(\frac{1}{2\widetilde{\rho}^2_b})$ vary from codeword to codeword.

Based on this understanding, we make the following modifications to our previous design to shape the final architecture of AttentionCode. Define $\overline{\bm{w}}^\prime\triangleq\bm{w}+\bm{w}^\prime$.
\begin{itemize}
\item In the first phase, we use QPSK instead of BPSK modulation.  In $K/2$ interactions, node A observes $K/2$ complex noise realizations from the feedback channel, and we denote them by $\overline{\bm{w}}^\prime_{0}=\{\overline{w}^\prime_{0,\ell}+j\overline{w}^\prime_{0,\ell+1}:\ell=1,3,5,...,K-1\}$. Note that $\overline{w}^\prime_{0,\ell}$ and $\overline{w}^\prime_{0,\ell+1}$ are the noise realizations encountered by two consecutive source bits.
\item In the second phase, node A performs $K$ encodings. In the $k$-th encoding, node A transmits a complex symbol ($2$ real coded symbols) to node B and receives a complex feedback symbol. Let the aggregated noise realization observed by node A be $\overline{w}^\prime_{1,k}+j\overline{w}^\prime_{2,k}$. We modify the input matrix to the encoder/AttentionNet, i.e., $\widetilde{\bm{S}}(k)$ in \eqref{eq:input2}, to
\begin{eqnarray}
\overline{\bm{S}}(k)=\begin{bmatrix}
b_1 & b_2 & b_3 & ... & b_k \\ 
\overline{w}^\prime_{0,1} & \overline{w}^\prime_{0,2} & \overline{w}^\prime_{0,3} & ... & \overline{w}^\prime_{0,k}\\ 
\overline{w}^\prime_{1,1} & \overline{w}^\prime_{1,2}  & \overline{w}^\prime_{1,3} & ... & 0 \\ 
\overline{w}^\prime_{2,1} & \overline{w}^\prime_{2,2} & \overline{w}^\prime_{2,3} & ... & 0
\end{bmatrix}.
\end{eqnarray}
\item Since fading leads to varying feedforward and feedback noise powers, AttentionCode faces a wide range of noise powers/distributions during training. In this light, it is essential to provide the CSI to AttentionCode so that it is adaptive to various noise powers. As shown in Fig.~\ref{fig:AttentionNet}, CSI information is provided to the encoder after passing through a linear layer. In particular, the CSI information contains the following features
\begin{equation}
    \text{CSI} = \left[h^{\mathfrak{r}}, h^{\mathfrak{i}}, h^{\prime\mathfrak{r}}, h^{\prime\mathfrak{i}}, \frac{2\widetilde{\sigma}^2_f}{\lvert h \rvert^2}, \frac{2\widetilde{\sigma}^2_b}{\lvert h \rvert^2\lvert h^\prime \rvert^2} \right]^\top.
\end{equation}
where the first four values are the real and imaginary parts of channel coefficients $h$ and $h^\prime$, and the last two values are the feedforward and feedback noise powers in a block of transmission.
\end{itemize}

\begin{rem}[Unknown fading coefficients]
In AttentionCode, we have assumed that the fading coefficients $h$ and $h^\prime$ are known to node A to perform encoding. This can be easily achieved by sending a pilot before the codeword transmission.
On the other hand, if the CSI is unknown, it is still possible to use AttentionCode. The reason is that the coded symbols transmitted from node A to node B and fed back to node A can serve as pilots, from which the fading coefficients $h$ and $h^\prime$ can be estimated at node A.

Specifically, given the received vector $\widetilde{\bm{y}}^\prime$, node A can estimate $h$ and $h^\prime$ by maximizing their posterior probabilities, yielding
\begin{equation}
\widehat{h},\widehat{h}^\prime =
\arg\max_{h,h^\prime} f(\widetilde{\bm{y}}^\prime\mid h,h^\prime,\widetilde{\bm{x}}) f(h,h^\prime),
\end{equation}
where
\begin{eqnarray}
\hspace{-0.1cm}&& f(\widetilde{\bm{y}}^\prime\mid h,h^\prime,\widetilde{\bm{x}}) \propto 
 \exp\left(-\frac{\lvert\widetilde{\bm{y}}^\prime - h^\prime h\widetilde{\bm{x}}\rvert^2}{4(\lvert h^\prime \rvert^2 \widetilde{\sigma}^2_f+\widetilde{\sigma}^2_b)}  \right), \\
\hspace{-0.1cm}&& f(h,h^\prime) \propto
\exp\left(-\frac{\lvert h \rvert^2}{2\widetilde{\rho}^2_f}\right)
\exp\left(-\frac{\lvert h^\prime \rvert^2}{2\widetilde{\rho}^2_b}\right).
\end{eqnarray}
It is worth noting that, estimating $h$ and $h^\prime$ from $\widetilde{\bm{y}}^\prime$ does not bring any new information. Instead, it is a process of feature extraction from the received signal to enable efficient learning.
\end{rem}

In addition to the architectural innovations discussed in this section, we borrow the ideas of zero padding and power reallocation from DeepCode in AttentionCode. The complete architecture of AttentionCode is illustrated in Fig.~\ref{fig:innovations}.

\section{AttentionCode: Training Methods}\label{sec:V}
In wireless communications, the goal is to transmit a stream of bits reliably to the receiver.
When the DL approach is utilized for end-to-end communication system modelling, the way that the training dataset is generated is exactly the same as that the test dataset is generated.
Said in another way, the training and test datasets share the same distribution; it is possible that the test examples have already been seen in the training phase.
In the extreme case, for example, we can enumerate all bitstream patterns in the training dataset, such that all bitstreams to be tested have been seen in the training phase.\footnote{It is worth noting that our discussion does not include the emerging semantic communication applications \cite{bourtsoulatze2019deep,semanticTheory} that employ joint source-channel coding, in which case the source is not necessarily limited to bitstreams. Therefore, enumerating possible source signals may no longer be possible. The training and test datasets must be independent to ensure that the learned DL models generalize well and do not overfit the training dataset, as classical machine learning.}
We emphasize that this is fundamentally different from classical machine learning tasks, where the training and test datasets must be independent.
In a way, {\it overfitting} is also important in our problem as the training dataset is exactly the test dataset.
In the design of AttentionNet, for example, we do not employ any overfitting prevention techniques, such as dropout, which is widely used in different variations of transformer, as they hinder AttentionCode from achieving ultra-low BLER.

In the above context, the training of error correction codes is different from classical DL models.
In this section, we present several methodologies that we have observed to help in training AttentionCode. They are potentially beneficial to other DL applications in the area of wireless communications.

\subsection{Large-batch training}\label{sec:IVA}
AttentionCode is essentially an autoencoder, where wireless channel is modeled by one layer of the autoencoder. The training of AttentionCode aims to minimize the binary cross-entropy loss in \eqref{eq:loss} using mini-batch stochastic gradient.

Unlike other machine learning tasks such as image recognition or natural language processing that target a misclassification rate of $10^{-2}\sim 10^{-3}$,
AttentionCode targets ultra-reliable communications with a BLER lower than  $10^{-5}$.
At such a low BLER, most mini-batches cannot observe a single error at a moderate SNR regime.
As a result, most mini-batches yield zero gradients if a small mini-batch size is used, and the training can easily converge to a suboptimum.

In DeepCode, the authors used a small mini-batch size of $B=200$ in training. Their results indicated that DeepCode cannot learn a well-structured code when the training input sequence is too short (e.g., $K=50$). In the experiments, the DNN was trained under an input sequence length $K=100$ and tested for an input sequence length $K=50$.

In contrast, AttentionCode uses a large mini-batch size in training (up to $B=10^4$) to eliminate the training instabilities caused by small mini-batches. We empirically found that, with large-batch training,
i) much better  feedback codes than DeepCode can be learned;
and ii) well-structured codes can be directly learned  even with a small $K$.
In general, we believe large-batch training will prove useful technique for general DL-based wireless communications applications that target a low BLER. 

\subsection{Distributed learning}
Input restructuring and large-batch training are very useful schemes for the feedback-code design. The downside, however, is that they lead to a much larger computation graph inside GPU in training and consume far more GPU resources than the most advanced GPU can provide.

To address this problem, AttentionCode trains the model in a distributed manner. Specifically, suppose all the parameters of AttentionCode form a vector $\bm{u}$. In the $t$-th training, we update $\bm{u}(t)$ to $\bm{u}(t+1)$ by sampling a batch of $B$ bit streams $\bm{b}(j)$, $j=1,2,...,B$, and $\bm{u}(t+1)$ is given by
\begin{eqnarray}\label{eq:distributed1}
\bm{u}(t+1)\hspace{-0.2cm}&=&\hspace{-0.2cm}\bm{u}(t) - \alpha(t) \sum_{j=1}^{B} \frac{\partial\mathcal{L}(\bm{b}(j),\bm{u}(t)) }{\partial \bm{u}(t)} \nonumber\\
\hspace{-0.2cm}&\triangleq&\hspace{-0.2cm} \bm{u}(t) - \alpha(t) \bm{g}\big(\bm{u}(t),B\big),
\end{eqnarray}
where $\alpha(t)$ is the learning rate and $\bm{g}\big(\bm{u}(t),B\big)$ is the gradient computed from $B$ batches. Eq.~\eqref{eq:distributed1} can be further written as
\begin{eqnarray}\label{eq:distributed2}
\bm{u}(t+1)\hspace{-0.2cm}&=&\hspace{-0.2cm} \bm{u}(t) - \alpha(t) \bm{g}\big(\bm{u}(t),B\big)\nonumber\\
\hspace{-0.2cm}&=&\hspace{-0.2cm} \bm{p}(t) - \alpha(t)\sum_{v=1}^{V} \bm{g}\big(\bm{u}(t),B_v\big).
\end{eqnarray}
That is, if we want to train a large batch size $B$, it is equivalent to
\begin{enumerate}[i)]
\item partition $B$ batches to $V$ parts of size $B_v$, $v=1,2,...,V$, where $B=\sum_{v=1}^V B_v$;
\item update $\bm{u}(t)$ by each smaller part and obtain the gradient $\bm{g}\big(\bm{u}(t),B_v\big)$;
\item update the DNN following \eqref{eq:distributed2}.
\end{enumerate}

In this way, the batch size for each training component is only $B_v$. Depending on the GPU capacity, $B_v$ can be as small as possible and the large training model is no longer a problem. We emphasize that distributed learning can be implemented over multiple GPUs or a single GPU in a time-division fashion. It is a useful technique in general when one has few GPU resources but wants to train with a particularly large batch.

\begin{rem}[Advanced optimizers]
The update rule of \eqref{eq:distributed1} is derived for the basic optimizer that performs mini-batch gradient descent. In this case, distributed learning in \eqref{eq:distributed2} is equivalent to \eqref{eq:distributed1}. On the other hand, advanced optimizers, such as Adam or AdaMax \cite{Adam}, often take the momentum of gradients and adaptive learning rates into account. In that case, distributed training and direct training with a large batch are no longer mathematically equivalent. Nevertheless, our empirical results validate that their performances are on an equal footing.
One can still use distributed training to mimic large-batch training when advanced optimizers are used.
\end{rem}

\subsection{Look-ahead optimizer}
AttentionCode uses the Adam optimizer
\cite{Adam} in conjunction with a look-ahead scheme to compute the gradient and update the parameters. Specifically, given AttentionCode with parameters $\bm{u}(t)$ and a batch size $B$,
\begin{enumerate}[i)]
    \item we first use the Adam optimizer to update $\bm{u}(t)$ for $C$ steps. That is, for $c=1,2,...,C,$,  we have
\begin{equation}\label{eq:lookahead1}
\bm{u}^{(c)}(t)= \bm{u}^{(c-1)}(t) - \bm{g}^{(c-1)}\big(\bm{u}^{(c-1)}(t),B\big),
\end{equation}
where $\bm{u}^{(0)}(t)=\bm{u}(t)$ and $\bm{g}^{(c-1)}\big(\bm{u}^{(c-1)}(t),B\big)$ is the scaled gradient direction (scaled by an adaptive learning rate) computed by the Adam optimizer.
\item then, the real update direction is $\bm{u}^{(C)}(t)-\bm{u}^{(0)}(t)$ and the updated AttentionCode $\bm{u}(t+1)$ is given by
\begin{equation}\label{eq:lookahead2}
\bm{u}(t+1)=\bm{u}(t) + \frac{1}{C}\Big(\bm{u}^{(C)}(t)-\bm{u}^{(0)}(t)\Big).
\end{equation}
\end{enumerate}

\begin{figure}[t]
    \centering
    \includegraphics[width=1\columnwidth]{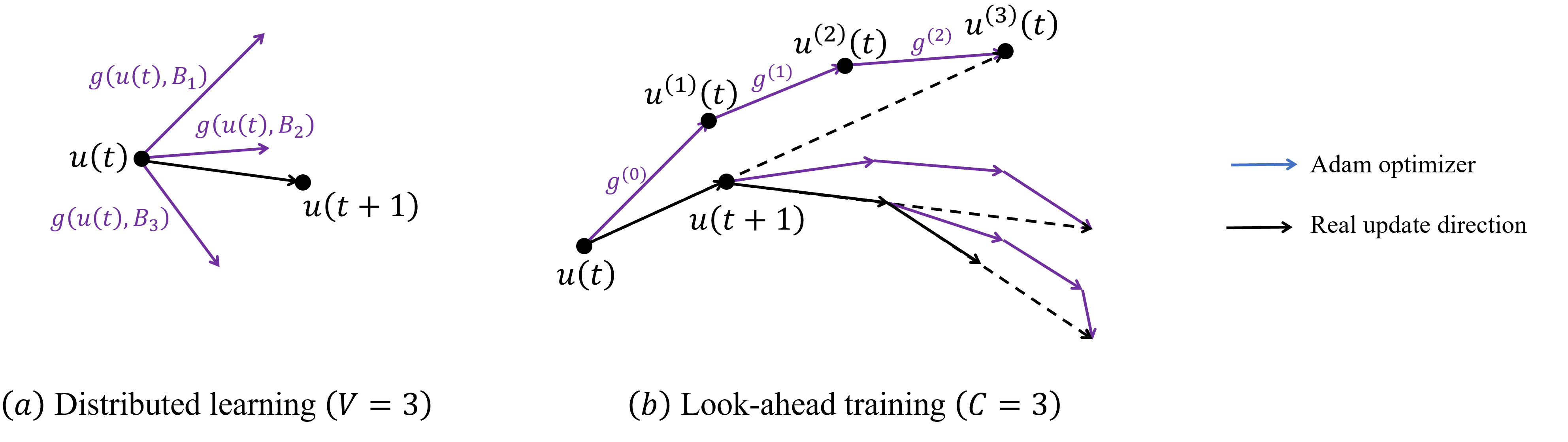}
    \caption{(a). Distributed learning: the real update direction is an average of $V$  independent gradients learned from $V$ mini-batches of sizes $B_v$; (b). Look-ahead optimizer: the real update direction is an average of $C$ gradients obtained from $C$ consecutive training steps, each  step is trained with a large batch size $B$.}
    \label{fig:lookahead}
\end{figure}

Fig.~\ref{fig:lookahead}(b) illustrates the look-ahead processes in \eqref{eq:lookahead1} and \eqref{eq:lookahead2}. As can be seen, the Adam optimizer is used to train $\bm{u}(t)$ for $C$ consecutive steps, while the real update direction is along $\bm{u}^{(C)}(t)-\bm{u}^{(0)}(t)$. Fig.~\ref{fig:lookahead} also highlights the difference between distributed learning and look-ahead training. These two schemes are orthogonal to each other, each Adam training step in Fig.~\ref{fig:lookahead}(b) can be implemented by distributed learning if a large batch size $B$ is used.

\begin{rem}[Look-ahead optimizer]
Our look-ahead scheme falls into the scope of the look-ahead optimizer proposed in \cite{lookahead}. The look-ahead optimizer consists of an inner optimizer and an outer optimizer.
Mapping to our approach, the Adam optimizer is the inner optimizer, and the average process in \eqref{eq:lookahead2} is the outer optimizer with a constant learning rate $1/C$.
\end{rem}

\subsection{Training-test SNR mismatch}\label{sec:mismatch}
In AttentionCode, the encoder learns to create correlations among coded symbols to combat channel impairments; the decoder, on the other hand, treats the decoding process as a classification problem -- a low BLER can be achieved provided that the optimal decision boundaries among all possible transmitted bitstreams are learned.
Imagine each bitstream as a point in a $K$ dimensional space. When the SNR is high, the decision boundaries are easier to determine since the ``noise balls'' are far from each other. On the other hand, when the SNR is low, the decision boundaries are harder to determine since the ``noise balls'' overlap with each other. 

Let there be two DNN-based decoders trained at a high SNR and a low SNR, respectively.
\begin{itemize}
\item When tested at the low SNR, the decoder trained in the high SNR may not work well, because its decision boundaries do not need to be finely crafted to achieve a low training loss in the high-SNR regime.
\item When tested at the high SNR, the decoder trained in the low-SNR regime can work to some extent, because its decision boundaries are learned to tackle much larger noise.
\end{itemize}

In light of the above, we propose to exploit  the ``training-test SNR mismatch'' to boost the performance of the learned feedback codes. 
In AttentionCode, there are two sorts of training-test SNR mismatches. Specifically, for a given test SNR, 
\begin{enumerate}
\item AttentionCode can be trained at a slightly lower SNR. For example, if the test SNR is $1$ dB, then we train AttentionCode at $0.8$ dB.
\item AttentionCode can be first trained at a lower SNR, and then retrained at the test SNR. For example, if the test SNR is $1$ dB, we first train AttentionCode at $0$ dB and retrain the trained model at $1$ dB.
\end{enumerate}

\subsection{Curriculum learning}
When the feedforward and feedback channels are very noisy, it is often not easy for AttentionCode to learn well-structured feedback codes from scratch. To address this problem, we leverage curriculum learning \cite{Curriculum} to train AttentionCode when the feedforward and feedback channel SNRs are low.

Curriculum learning refers to a type of learning method where the training starts out with easy tasks and then gradually faces more and more difficult tasks. As stated in Section~\ref{sec:mismatch}, the training of AttentionCode is essentially a classification problem and the difficulty of classification depends on the SNR. Therefore, when the SNR is low and direct learning does not work out well, we can start with high-SNR examples and gradually decrease the SNR of the training examples so that the DNN  faces gradually more difficult tasks. This scheme is observed to be particularly beneficial to noisy feedback channels.

\section{Numerical Experiments}\label{sec:VI}
To verify the effectiveness of the architectural innovations in Section \ref{sec:IV} and training innovations in Section \ref{sec:V}, this section performs numerical experiments to evaluate the  performance of AttentionCode benchmarked against prior solutions. 
For a fair comparison, we consider the same experimental setup as \cite{DeepCode,DEFC,DRFC}, where the bitstream length is $K=50$ and the code rate $R=1/3$. The feedback channel can be noiseless and noisy ($\eta^\prime=20$ dB).

\begin{table}[t]
\renewcommand*{\arraystretch}{0.6}
\centering
\caption{Hyper-parameter settings for AttentionCode}
\begin{tabular}{@{}ccc@{}}
\toprule
Parameters & Descriptions              & Values   \\
\midrule
$K$       & bitstream length & $50$        \\
$R$       & code rate & $1/3$        \\
$d_S$       & input dim & $4$ (Enc), $3$ (Dec)        \\
$q_t$       & \#enc blocks in the encoder & $2$        \\
$q_r$       & \#enc blocks in the decoder & $3$        \\
$d_m$       & model size & $32$        \\
$B$         & batch size   & $2000$       \\
$V$         & distributed learning factor   & $5$       \\
$C$         & look-ahead steps   & $4$       \\
Adam            & inner optimizer   & $\beta_1\!=\!0.9, \beta_2\!=\!0.98, \epsilon\!=\!{10}^{-9}$       \\
$\text{lr}$ & learning rate   & $10^{-3}$        \\
\bottomrule
\end{tabular}
\label{tab:parameter}
\end{table}

For AttentionCode, the hyper-parameter settings are summarized in Table \ref{tab:parameter}.
For AttentionNet, we set the number of encoding blocks at the transmitter and receiver to $q_t=2$ and $q_t=3$, respectively. We have observed that stacking more encoding blocks degrades the performance.
The model size $d_m$ is set to $32$, which is a good balance between performance and complexity.
Multi-head attention and dropout are not incorporated, as discussed in Section \ref{sec:IV}.
For training, we use both distributed learning and the look-ahead mechanism. Unless specified otherwise, we set the batch size to $B=2000$, the distributed factor to  $V=5$, and the look-ahead steps to $C=4$.

\subsubsection{Ablation study}
In the first experiment, we perform an ablation study to verify the effectiveness of different components of AttentionCode. Specifically, we consider the following four models:
\begin{itemize}
\item AttentionCode: the complete version of AttentionCode.
\item AttentionCode-M : remove the mask from AttentionCode.
\item AttentionCode-M-IR: remove the input restructuring and mask from AttentionCode.
\item AttentionCode-TT-C: remove train-test SNR mismatch and curriculum learning from AttentionCode.
\end{itemize}


\begin{figure}[t]
\centering
\subfigure{
\includegraphics[width=0.75\linewidth]{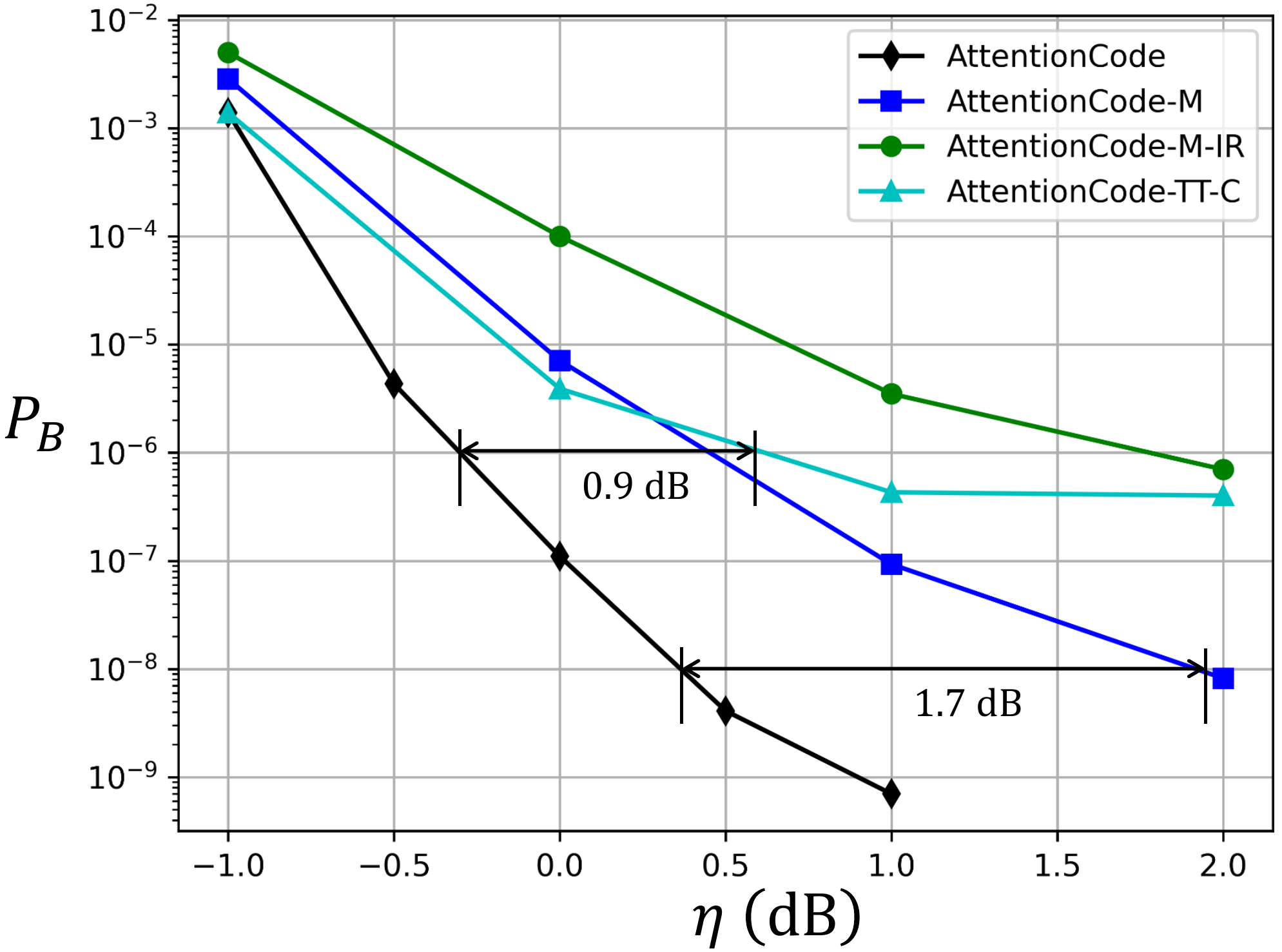}
}
\caption{An ablation study to evaluate the effectiveness of different modules in AttentionCode. The BLER  performance $P_B$ versus feedforward channel SNR $\eta_r=\eta$ (in dB) in AWGN channels. The feedback is noiseless.}
\label{fig:S0}
\end{figure}

The comparisons among the four models are presented in Fig.~\ref{fig:S0}, where we consider the AWGN-channel case with a noiseless feedback channel and evaluate the BLER $P_B$ versus the feedforward channel SNR $\eta$ (in dB) performance. As can be seen, input restructuring and mask are very important to AttentionCode. Without a mask, the performance of AttentionCode degrades by $1.7$ dB to achieve a BLER of $10^{-8}$. Without input restructuring, the performance further deteriorates.
On the other hand, train-test SNR mismatch
and curriculum learning are essential to AttentionCode to achieve low BLER at the high-SNR regime -- without them, the BLER performance degrades for $0.9$ dB to achieve a BLER of $10^{-6}$ and exhibits an error floor at the high-SNR regime.

Next, we explain in more detail how the best performance of AttentionCode in Fig.~\ref{fig:S0} is achieved. First, a different model is trained for each forward SNR $\eta$.
Second, the training processes consist of two steps.
\begin{enumerate}[i)]
\item Curriculum learning: The training starts from $\eta=1$ dB. We train AttentionCode from scratch at $\eta=1$ dB and obtain an initial model, denoted by $\bm{u}(i,1)$, when the training converges.
Then, we decrease the training SNR $\eta$ from $1$ to $-1$ (decreased by $0.5$ each time) and retrain $\bm{u}(i,\eta)$ progressively (note that $\bm{u}(i,\eta)$ is trained from $\bm{u}(i,\eta+0.5)$ instead of $\bm{u}(i,1)$). Eventually, we attain an AttentionCode models at $\eta=-1$ dB, i.e., $\bm{u}(i,-1)$.
Over the course of curriculum learning, the batch size, distributed factor, and look-ahead step are set according to Table~\ref{tab:parameter}.
\item Training-test SNR mismatch: We next use the training-test SNR mismatch to obtain the AttentionCode model at each SNR. In particular, we start from $\eta=-1$ dB and progressively increase the SNR to $\eta=1$ dB. In this process, the model $\bm{u}(i,\eta)$ is obtained from $\bm{u}(i,\eta-0.5)$ by either setting $\bm{u}(i,\eta)=\bm{u}(i,\eta-0.5)$ or retrain $\bm{u}(i,\eta-0.5)$ at $\eta$, as stated in Section~\ref{sec:mismatch}. If retraining is needed, the look-ahead step is set to $C=1$, while the batch size and the distributed learning factor remain the same.
\end{enumerate}

\subsubsection{AWGN channels}
Next, we compare the performance of AttentionCode with all existing schem\-es, assuming both feedforward and feedback channels are AWGN channels.

\begin{figure*}[t]
\centering
\subfigure{
\includegraphics[width=0.38\linewidth]{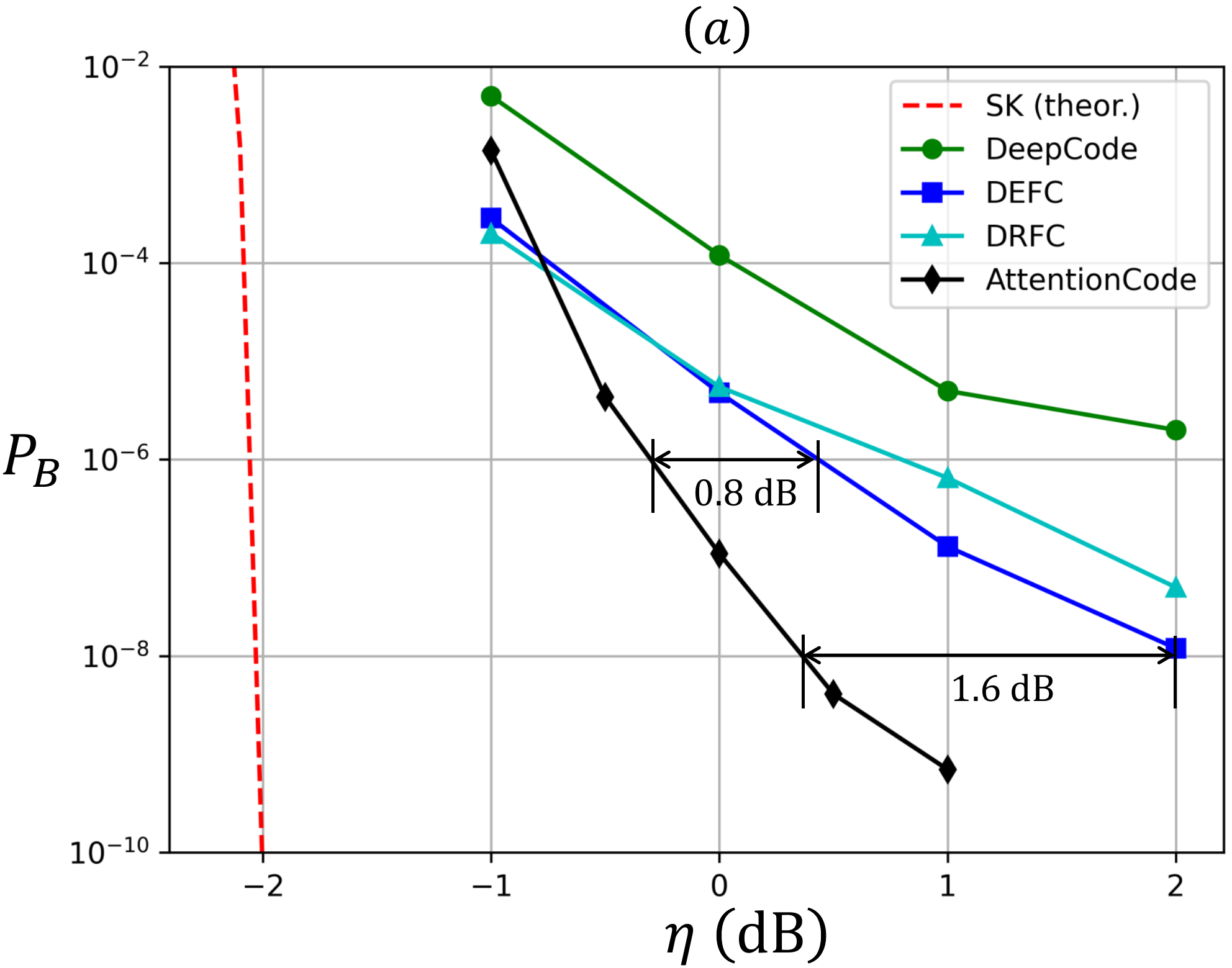}
}
\quad\quad
\subfigure{
\includegraphics[width=0.38\linewidth]{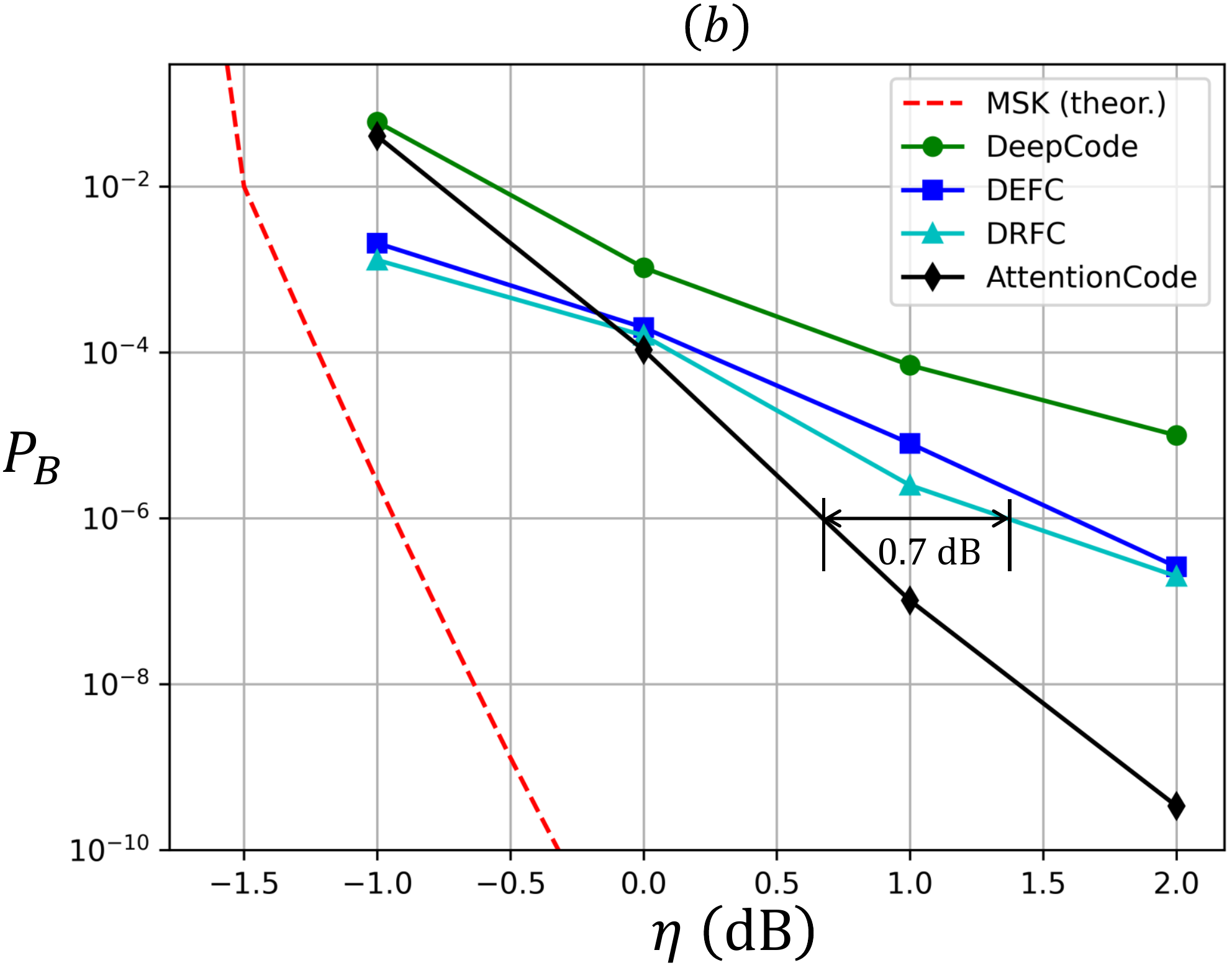}
}
\caption{AWGN channels: BLER $P_B$ versus feedforward channel SNR $\eta$ (in dB) performance of various feedback codes. (a) noiseless feedback, where $\eta^\prime=\infty$; (b) noisy feedback, where $\eta^\prime=20$ dB.}
\label{fig:S1}
\end{figure*}

\begin{figure*}[t]
\centering
\subfigure{
\includegraphics[width=0.35\linewidth]{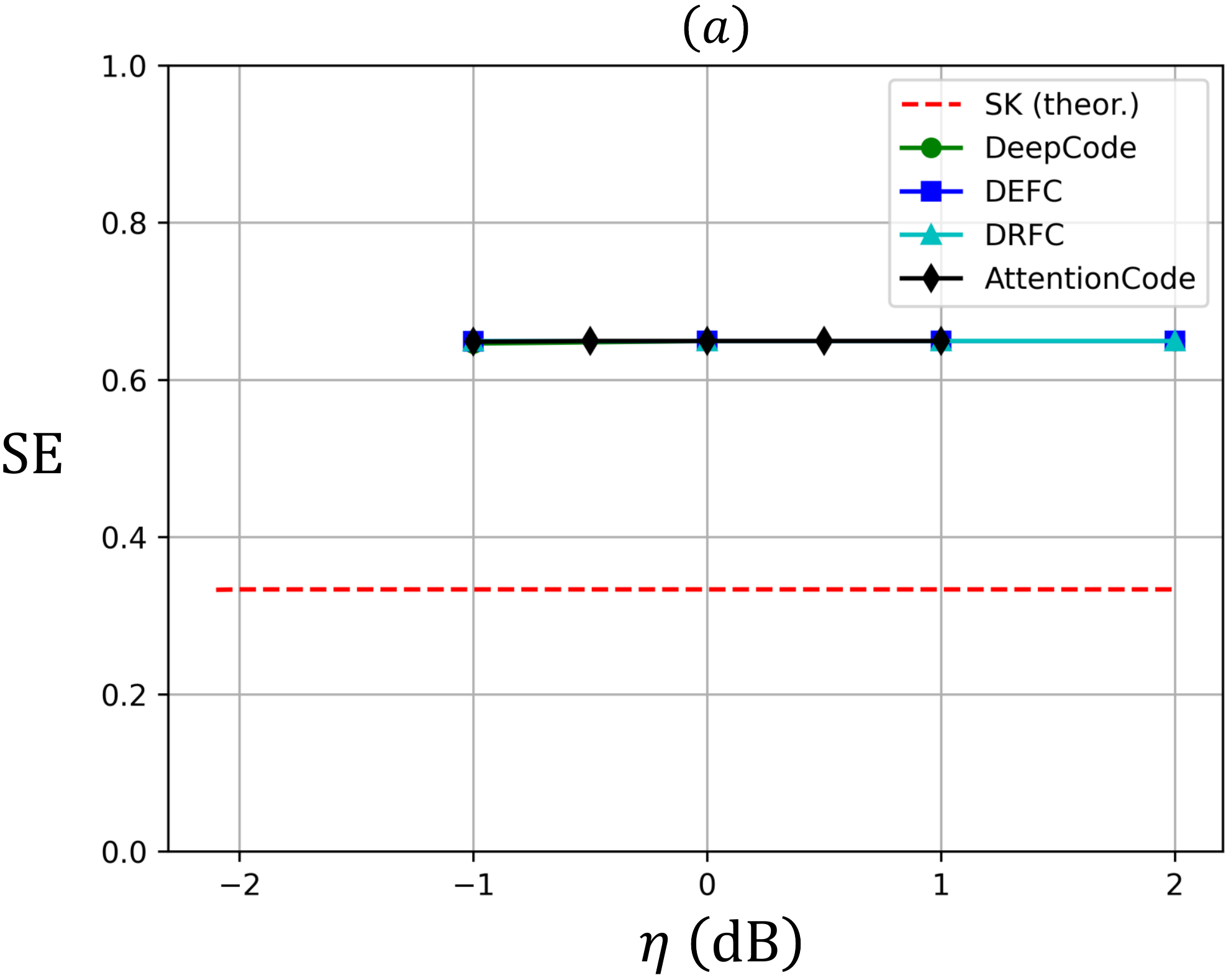}
}
\quad\quad\quad
\subfigure{
\includegraphics[width=0.35\linewidth]{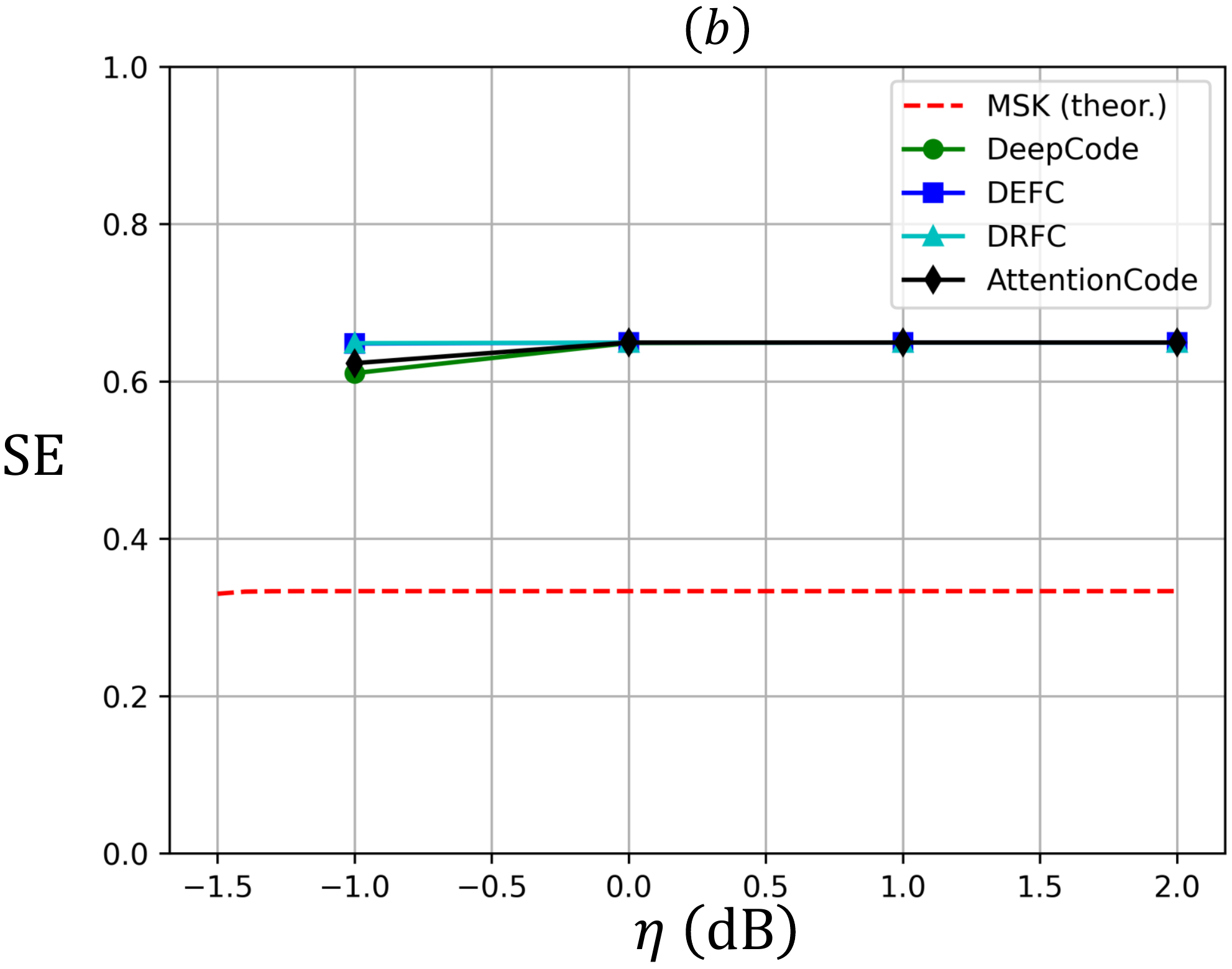}
}
\caption{AWGN channels: spectral efficiency versus feedforward channel SNR $\eta$ (in dB) performance of various feedback codes. The code rate is fixed to $1/3$. (a) noiseless feedback, where $\eta^\prime=\infty$; (b) noisy feedback, where $\eta^\prime=20$ dB.}
\label{fig:S2}
\end{figure*}


\begin{figure*}[t]
\centering
\subfigure{
\includegraphics[width=0.38\linewidth]{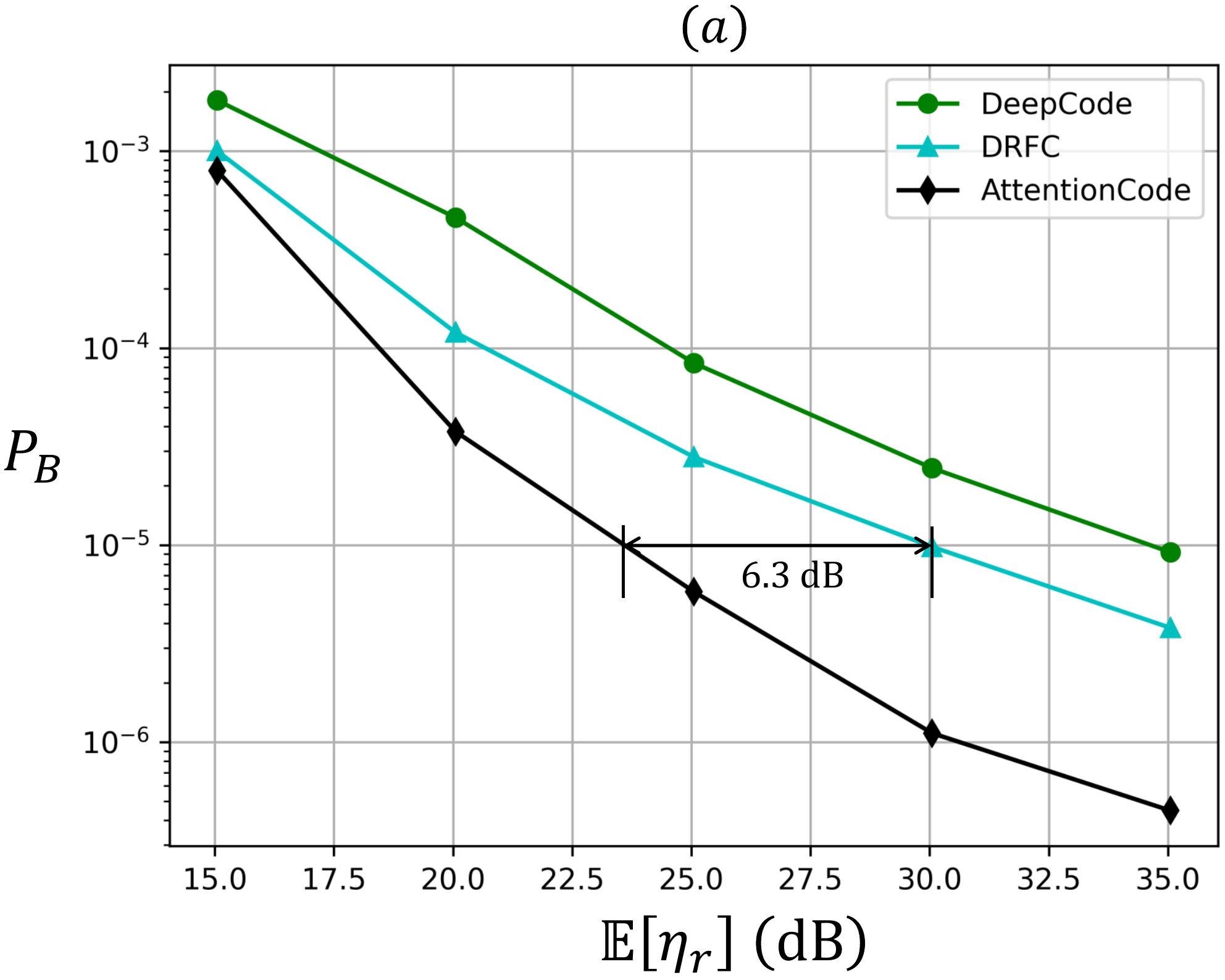}
}
\quad\quad
\subfigure{
\includegraphics[width=0.38\linewidth]{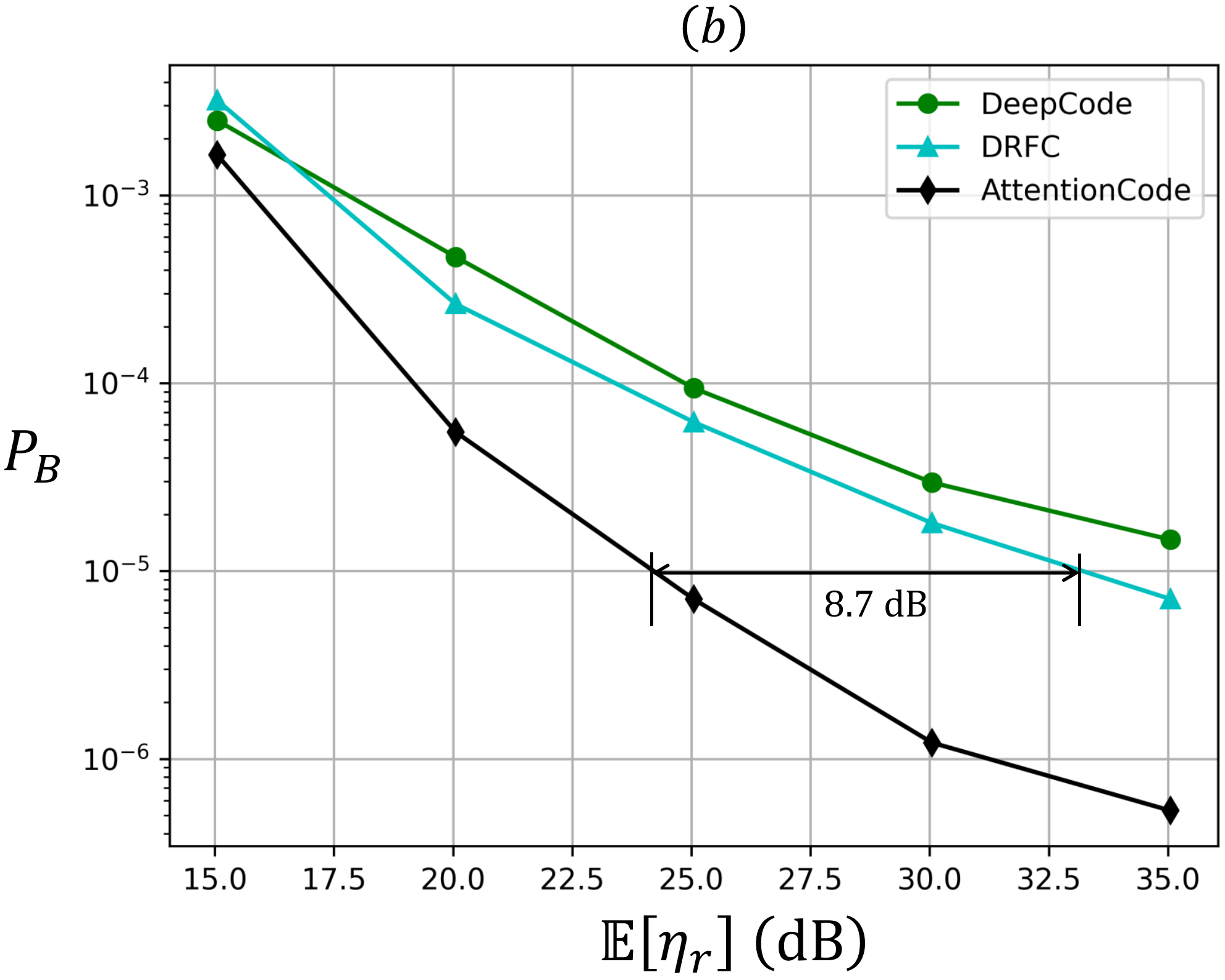}
}
\caption{Fading channels: BLER $P_B$ versus the average received SNR $\mathbb{E}[\eta_r]$ (in dB) performance of various feedback codes. (a) the feedforward channel is a fading channel with $\widetilde{\rho}_f=4$; the feedback channel is noiseless; (b) both feedforward and feedback channels are fading channels with $\widetilde{\rho}_f=4$ and $\widetilde{\rho}_b=1$. For the feedback channel, we fix $\eta^\prime=20$ dB.}
\label{fig:S3}
\end{figure*}

\begin{figure*}[t]
\centering
\subfigure{
\includegraphics[width=0.38\linewidth]{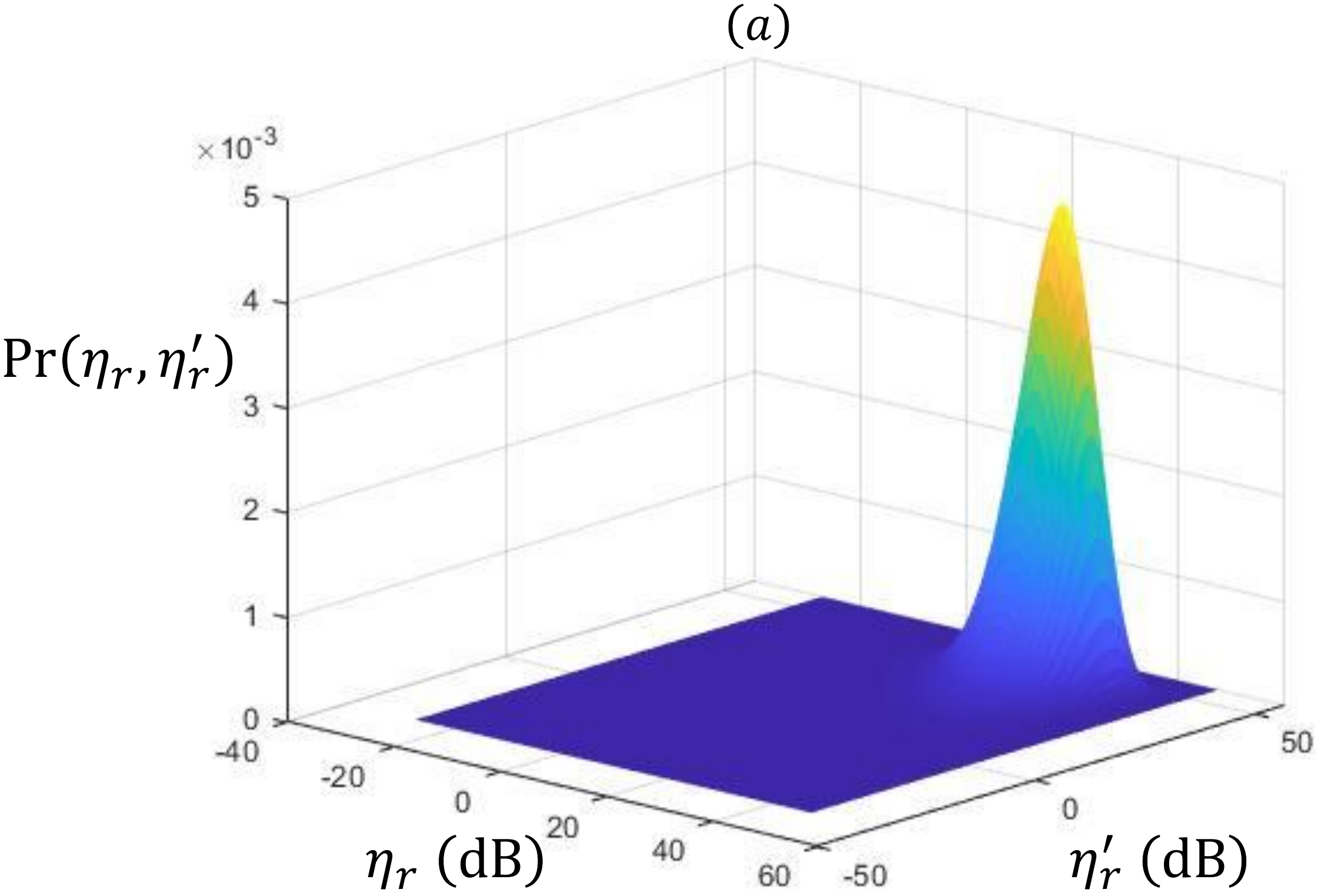}
}
\quad\quad
\subfigure{
\includegraphics[width=0.35\linewidth]{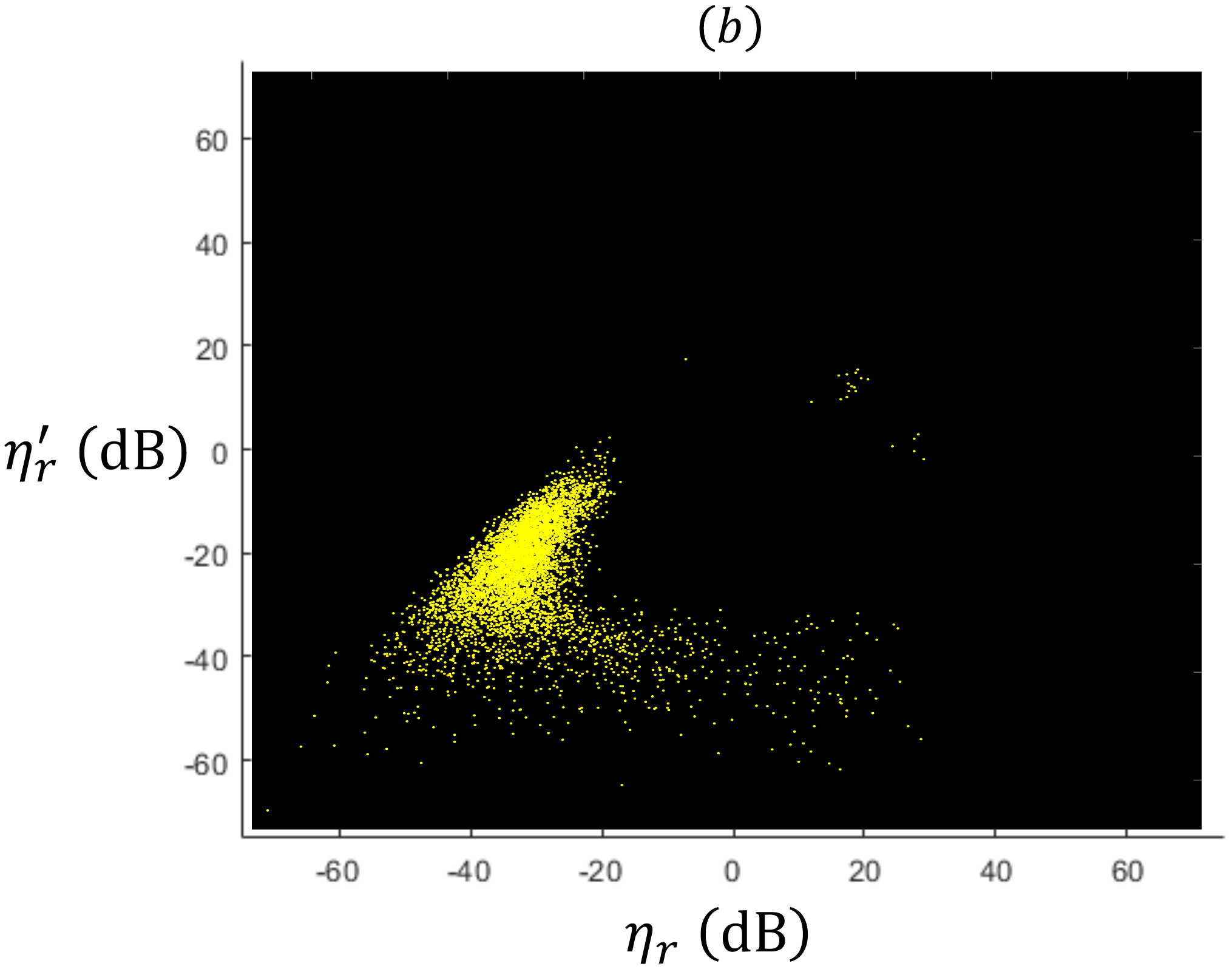}
}
\caption{(a) The joint probability density function of $\eta_r$ and $\eta^\prime_r$, i.e., the SNRs of the feedforward and feedback channels, wherein $\widetilde{\rho}_f=4$, $\widetilde{\rho}_b=1$, $\eta=20$ dB, and $\eta^\prime=20$ dB; (b) distribution of errors on the instantaneous-SNR plane.}
\label{fig:S4}
\end{figure*}

Fig.~\ref{fig:S1} presents the BLER performance $P_B$ versus the feedforward channel SNR $\eta$ (in dB) for the SK scheme, the modulo-SK scheme, DeepCode, DEFC, DRFC, and AttentionCode. In particular, Fig.~\ref{fig:S1}(a) consider a noiseless feedback channel and Fig.~\ref{fig:S1}(b) considers a noisy feedback channel with a SNR $\eta^\prime=20$ dB.

We first compare AttentionCode with existing DL-based feedback codes.
As can be seen, AttentionCode achieves a new state of the art in terms of the BLER performance.
In the noiseless feedback setup, AttentionCode outperforms the best existing code (i.e., DEFC) by $0.8$ dB to attain a BLER of $10^{-6}$, and by $1.6$ dB to attain a BLER of $10^{-8}$.
In the noisy feedback setup,  AttentionCode outperforms the best existing code (i.e., DRFC) by $0.7$ dB to attain a BLER of $10^{-6}$.

In terms of complexity, the training complexity of AttentionCode is much higher than RNN-based codes (i.e., DeepCode, DEFC, and DRFC) as it consumes more time and computation resources. Also, we need a combination of the training methods proposed in Section~\ref{sec:V}.
However, once the models are trained, one can reap larger gains from AttentionCode, as demonstrated in Fig.~\ref{fig:S1}, and the test complexity of AttentionCode (i.e., apply well-trained AttentionCode for transmission) is on an equal footing with RNN-based codes.

\begin{rem}[Computational complexity of AttentionCode]\label{rem:comp}
The training and execution complexities of AttentionCode and prior works can be compared in a more quantitative fashion.

Let us first consider the number of computations in the neural networks.
Consider a ($K$,$N$) feedback code. Existing DL-based feedback code designs, e.g., DeepCode, DEFS, and DRFS, leverage RNNs. The number of computations in the RNN is $\mathcal{O}\big(Kd^2_{\text{RNN}}\big)$, where $d_{\text{RNN}}$ is the size of latent variables, i.e., the number of latent features we use to represent one input. In prior works, the common settings are $K=50$ and $d_{\text{RNN}}=50$. 
AttentionCode, on the other hand, leverages the transformer with the self-attention mechanism. The number of computations in the self-attention mechanism is $\mathcal{O}\big(K^2d_{\text{Atten}}\big)$. In the experiments, we set $K=50$ and $d_{\text{Atten}}=32$.

The training complexity is proportional to both the number of computations in the neural networks and the number of training epochs $T$. Thus, the training complexities of prior works and AttentionCode are $\mathcal{O}\big(T_{\text{RNN}}Kd^2_{\text{RNN}}\big)$ and $\mathcal{O}\big(T_{\text{Atten}}K^2d_{\text{Atten}}\big)$, respectively. Since $T_{\text{Atten}}\gg T_{\text{RNN}}$, the training complexity of AttentionCode is much larger than prior works. Empirically, 1) if we set $T_{\text{Atten}}= T_{\text{RNN}}$, AttentionCode cannot outperform DeepCode; 2) if we set $T_{\text{RNN}}= T_{\text{Atten}}$, the performance of DeepCode does not improve further and is still much worse than AttentionCode.

However, it is worth noting that the training of feedback code is a one-shot cost. Once trained, the feedback code can be deployed and only the execution complexity matters. As stated above, the execution cost of existing feedback codes and AttentionCode are $\mathcal{O}\big(Kd^2_{\text{RNN}}\big)$ and $\mathcal{O}\big(K^2d_{\text{Atten}}\big)$, respectively. At a short block length (e.g., $K=50$), the execution complexity of AttentionCode is on an equal footing with existing codes.
\end{rem}

\begin{rem}[Forward error correction code]
As an additional benchmark to measure how much feedback improves the reliability of the feedforward channel, new radio (NR) LDPC, which does not exploit the feedback, achieves a BLER of $2\times 10^{-3}$ when the feedforward SNR is  $\eta=1$ dB; and a BLER of $3\times 10^{-5}$ when the feedforward SNR is $\eta=2$ dB.
\end{rem}

\begin{rem}[forward error correction code with doubled resources]
Another benchmark is forward error correction code with doubled channel resources. For this benchmark, the total channel resources consumed by feedforward and feedback links are the same as AttentionCode.
Our results show that AttentionCode with $R=1/3$ cannot outperform LDPC with $R=1/6$. 
This, however, does not mean that doubling the feedforward channel resources is a better option in practice because the feedforward and feedback channel resources are often not interchangeable. Consider the uplink transmission in a slotted cellular system, where some of the time-frequency resources in a slot are assigned to uplink, and others are assigned to downlink.
\begin{itemize}
\item First, doubling the uplink resources requires a global resource reallocation. Feedback-aided communication, on the other hand, can directly leverage the downlink resources to improve the uplink reliability.
\item Second, doubling the uplink resources consumes additional energy of the mobile user, whose battery is very limited. In contrast, feedback-aided communication consumes the energy of the base station, which is infinite. 
\end{itemize}
\end{rem}

Next, we compare AttentionCode with the SK and modulo-SK schemes.
As detailed in Section~\ref{sec:II}, the SK and modulo-SK schemes are designed based on the idea of transmitting a $2^K$-PAM constellation from node A to node B, and successively refining the estimate of node B in the following interactions.
In each interaction, node B describes the current estimation error to node A via the feedback channel.
In Fig.~\ref{fig:S1}, we have presented their theoretical performance as benchmarks to DL-based codes in the noiseless and noisy feedback setups, respectively.
The practical performance of these two schemes, on the other hand, can degrade severely (worse than LDPC which does not exploit feedback) due to the precision and quantization issues stated in Section \ref{sec:II}.
More discussion and debate on the precision issues of the SK and modulo-SK schemes can be found in \cite{response1} and \cite{response2}.
In general, we believe the SK and modulo-SK schemes are good candidates when i) the bitstream length is short, e.g., $K\leq 16$; ii) both the feedforward and feedback channels are AWGN channels with fixed SNRs.

Another angle to compare the SK and modulo-SK schemes with DL-based feedback codes are the spectral efficiency, defined as the number of successfully transmitted bits per interaction, i.e.,
\begin{equation}
  \text{SE}=\frac{(1-P_B)K}{\widetilde{N}}.  
\end{equation}
As explained in Section~\ref{sec:II}, the SK and modulo-SK schemes are designed for unit-delay feedback. This suggests that, to transmit a codeword of length $N$, the SK and modulo-SK schemes require $\widetilde{N}=N$ interactions between nodes A and B.
In contrast, DL-based feedback codes are adaptive to longer feedback delays \cite{DeepCode}.
For the current design of DL-based codes (including DeepCode, DEFC, DRFC, and AttentionCode), the feedback delay is $2$ units, i.e., each interaction between nodes A and B carries $2$ real coded symbols. Thus, DL-based codes require $\widetilde{N}=N/2$ interactions to transmit a codeword of length $N$, yielding a higher spectral efficiency.

Fig.~\ref{fig:S2} compares the spectral efficiency of the SK and modulo-SK schemes and DL-based feedback codes, where $P_B$ are given by the experimental results in Fig.~\ref{fig:S1}; the bitstream length $K=50$; the number of interactions $\widetilde{N}=KR$ for the SK and modulo-SK schemes, and $\widetilde{N}= \lceil {(K+1)R}/{2} \rceil$ for DL-based codes (we have taken the extra spectrum consumed by zero-padding into account). As can be seen, DL-based codes almost double the spectral efficiency compared with the SK and modulo-SK schemes, thanks to their adaptability to longer feedback delays.

\subsubsection{Fading channels}
Finally, we evaluate the performance of AttentionCode in fading channels. Considering the fact that prior works focused on exclusively on AWGN channels (or real fading channels, such as DRFC), we first implement two prior schemes, DeepCode and DRFC, and assess their performances in fading channels as benchmarks. For a fair comparison, the information we feed into DeepCode and DRFC is exactly the same as AttentionCode, including source bits, noise realizations, CSI, etc. For DeepCode, we use GRUs as encoder and decoder with a batch size of $200$; for DRFC, we use LSTM as encoder and decoder with a batch size of $8192$.

In fading channels, the noise powers imposed by the feedforward and feedback channels on a transmitted symbol are $\frac{2\widetilde{\sigma}^2_f}{\lvert h \rvert^2}$ and $\frac{2\widetilde{\sigma}^2_b}{\lvert h \rvert^2\lvert h^\prime \rvert^2}$, respectively. 
The SNRs at nodes A and B are
$\eta_r= \lvert h \rvert^2\eta$ and
$\eta^\prime_r=\lvert h \rvert^2\lvert h^\prime \rvert^2\eta^\prime$.
In the experiments, we consider the SNRs of practical interest (BLER $\leq 10^{-5}$) by setting $\widetilde{\rho}_f=4$, $\widetilde{\rho}_b=1$, $\eta^\prime=20$ dB and varying $\eta$ in $[0,20]$ dB. Therefore, the average SNRs are $\mathbb{E}[\eta_r]= 2\widetilde{\rho}^2_f\eta \in [15.1,35.1]$ dB and $\mathbb{E}[\eta^\prime_r]= 4\widetilde{\rho}^2_f\widetilde{\rho}^2_f\eta^\prime\approx 38.1$ dB.

The performance comparisons among DeepCode, DRFC and AttentionCode are presented in Fig.~\ref{fig:S3}, wherein we have considered two fading scenarios. Specifically, in Fig.~\ref{fig:S3}(a), we consider a fading feedforward channel with  $\widetilde{\rho}_f=4$, while the feedback channel is noiseless. 
On the other hand, in Fig.~\ref{fig:S3}(b), both feedforward and feedback channels are fading, and we set $\widetilde{\rho}_f=4$, $\widetilde{\rho}_b=1$ and $\eta^\prime=20$ dB.
In both fading scenarios, AttentionCode achieves significant performance gains over DeepCode and DRFC. To attain a BLER of $10^{-5}$, AttentionCode outperforms DRFC by $6.3$ dB and $8.7$ dB, respectively, in two scenarios.

The BLER performance presented in Fig.~\ref{fig:S3} is the first-order statistic to characterize the block errors. In actuality, Rayleigh fading leads to large dynamic ranges of $\eta_r$ and $\eta^\prime_r$ and it is more revealing to measure the behavior of AttentionCode under each possible pair of instantaneous SNRs.
In this light, next we focus on the well-trained model at $\mathbb{E}[\eta_r]\approx 35.1$ dB in Fig.~\ref{fig:S3}(b) and measure which pair of instantaneous SNRs that leads to the most block errors.

First, we plot in Fig.~\ref{fig:S4}(a) the joint probability density function of $\eta_r$ and $\eta^\prime_r$ (in dB), considering the above fading setup.
As can be seen, the average SNRs are $\mathbb{E}[\eta_r]\approx 35.1$ dB and $\mathbb{E}[\eta^\prime_r]\approx 38.1$ dB.
In the worst case, both $\eta_r$ and $\eta^\prime_r$ can go to $-\infty$ dB. This poses challenges to DL as the learned models have to adapt to a wide range of SNR pairs, as opposed to only a fixed SNR pair in the AWGN-channel case.
Given the well-trained AttentionCode, we record the pairs of instantaneous SNRs that lead to a block error and plot them on the SNR plane, as shown in Fig.~\ref{fig:S4}(b). As can be seen, most errors are concentrated around the region $\{-45\leq\eta_r\leq -20, -40\leq\eta^\prime_r\leq 0\}$ (in dB), but the probability that the SNR pair falls into this region is quite small.

\section{Conclusion}\label{sec:Conclusion}
This paper put forth a new class of DL-aided channel codes, dubbed AttentionCode, that exploits the feedback from the receiver to enhance the reliability of feedforward transmissions. AttentionCode sets up a new state-of-the-art among DL-based feedback codes. The architectural and training innovations yield the following takeaway messages:
\begin{enumerate} 
\item The interaction-based communication paradigm is a pro\-mising solution to ultra-reliable short-packet communications for future wireless networks.
\item The success of AttentionCode validates the feasibility and superiority of pure-attention-based DNNs in the design of feedback codes. Further, the attention mechanism can be generalized to other channel coding applications as encoder and decoder to boost performance.
\item The design principles and training innovations proposed in this paper are potentially beneficial to general DL applications in wireless communications, where both generalization and overfitting are important in DNN training.
\end{enumerate} 

Moving forward, we point out two directions to explore along with the vision of this paper.
\begin{itemize}
\item Block feedback and unbalanced feedback. This paper focused on a setup of unit-time delay and balanced transmission, where each feedforward or feedback transmission delivers 1 complex symbol. In practice, the transmissions are packet-based, and one feedforward or feedback transmission can deliver a packet of symbols more than 1 -- this corresponds to a block feedback setup. On the other hand, the transmission resources of the transmitter and receiver (e.g., physical resource blocks (PRBs) and power budget) vary -- this corresponds to an unbalanced feedback setup. Both setups are of practical interest for further investigation.
\item Applications of the interaction-based communication par\-adigm. In existing multiple-input and multiple-output (MIMO) communication systems, CSI feedback is necessary for precoding, beamforming, and interference mitigation. With the interaction-based paradigm, the feedback link can be utilized not only for reliable channel coding but also for real-time CSI feedback. This raises a new research direction of joint channel-coding and beamforming with active feedback. More broadly, the interaction-based communication paradigm complements emerging topics of 6G, such as edge intelligence, semantic communications, and joint communication and sensing -- by exploiting the feedback link, significant performance gains for these systems are expected.
\end{itemize}

\appendices
\section{Additional simulations}\label{sec:AppA}
\begin{figure}[t]
\centering
\includegraphics[width=0.6\linewidth]{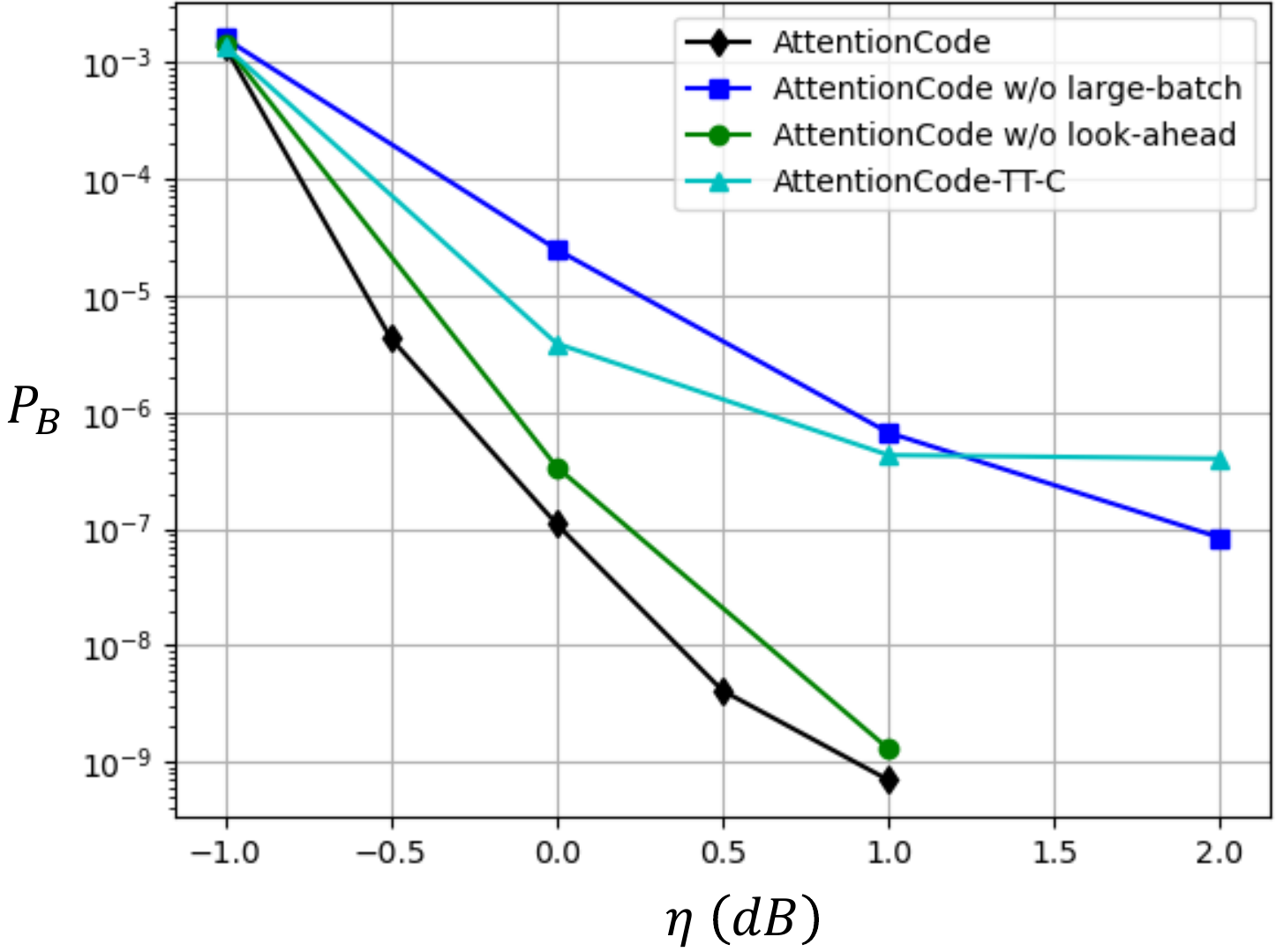}
\caption{Impact of different training methods on AttentionCode. The BLER performance $P_B$ versus feedforward channel SNR $\eta$ in AWGN channels. The feedback is noiseless.}
\label{fig:ablation}
\end{figure}

{\it Ablation study}:
we perform additional simulations to further evaluate the impacts of individual training methodologies on the BLER performance. It is worth noting that the effects of the training methods are not additive, instead, they often have to be used in conjunction with each other to be effective. In view of this, to evaluate the impact of a training method, we remove it from AttentionCode and evaluate the performance loss of AttentionCode.
The results are presented in Fig. \ref{fig:ablation}. As can be seen, large-batch training is essential for AttentionCode. Without it, the performance deteriorates by a significant margin. Distributed learning is a way to realize large-batch training, hence having the same impact on AttentionCode. The look-ahead optimizer, in comparison, has relatively less impact. Training-test SNR mismatch and curriculum learning are often used together. By removing them, AttentionCode does not perform well in the high-SNR regime.

{\it CSI error}:
CSI estimation error is inevitable in practice. To evaluate the robustness of AttentionCode to CSI errors, we assume that the CSI $(\widehat{h},\widehat{h}^\prime)$ available to the transmitter and receiver is a noisy version of the true CSI, i.e., $\widehat{h}=h+w_h$, $\widehat{h}^\prime=h^\prime+w^\prime_h$, where $w_h, w_h^\prime \sim\mathcal{CN}(0,2\widetilde{\rho}^2_e)$.

\begin{figure}[t]
\centering
\includegraphics[width=0.6\linewidth]{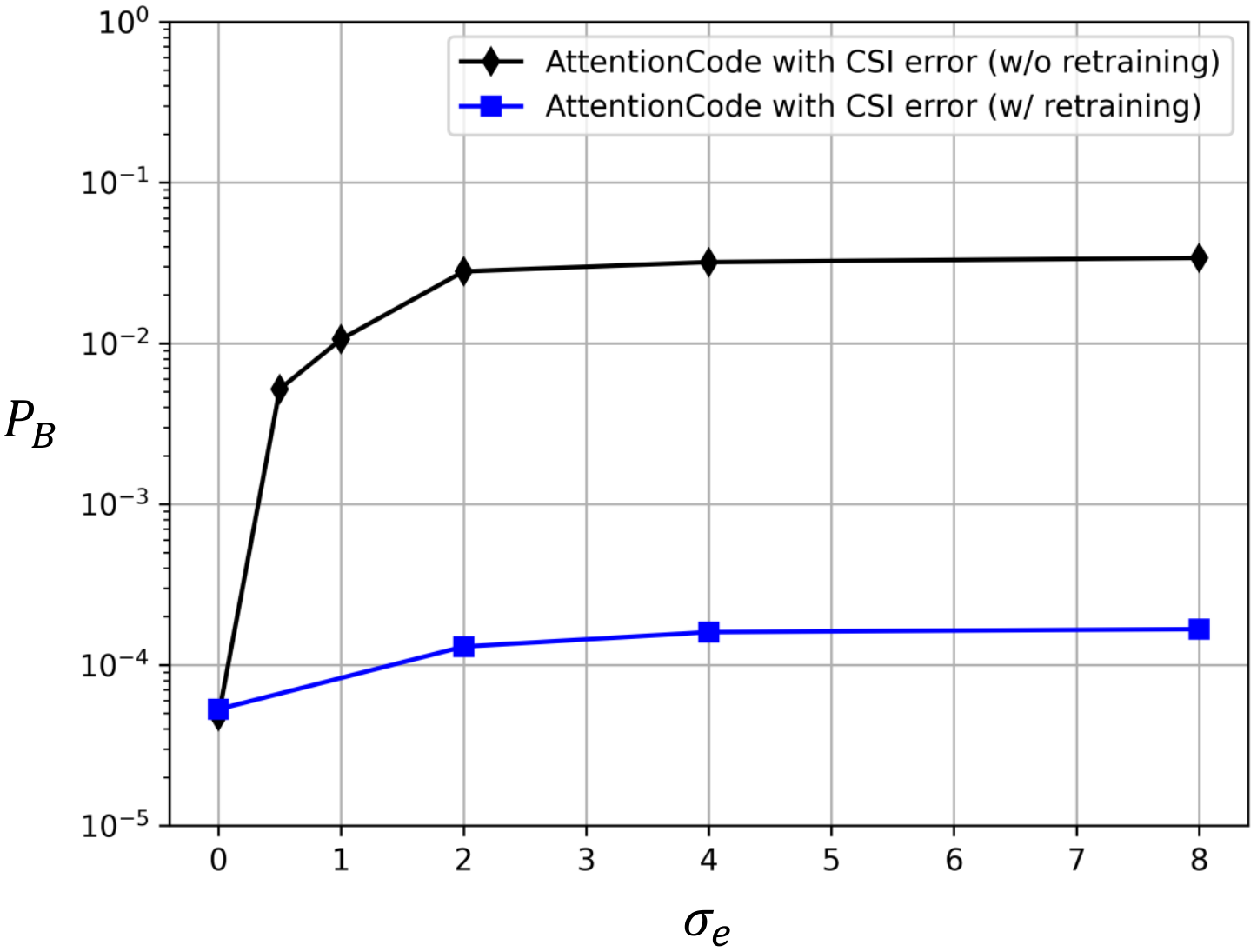}
\caption{Robustness of AttentionCode (without retraining and retrained with CSI error) to the inaccurate CSI.}
\label{fig:csi}
\end{figure}

Consider fading channels where the average SNRs of the feedforward and feedback channels are $\mathbb{E}[\eta_r]= 20$ dB and $\mathbb{E}[\eta^\prime_r]= 38.1$ dB, respectively. Given the noisy CSI, we evaluate the robustness of AttentionCode, which was well-trained without CSI error, to the CSI errors by varying $\sigma_e$ and present the simulation results in Fig. \ref{fig:csi} (the black curve). As shown, with the increase in $\sigma_e$, the BLER performance of AttentionCode quickly degrades. However, this can be efficiently mitigated by incorporating inaccurate CSI into the training phase. As can be seen from Fig. \ref{fig:csi} (the blue curve), AttentionCode retrained in the presence of inaccurate CSI is robust to CSI errors, the performance only degrades slightly even with a large estimation error.
 
\begin{figure}[t]
\centering
\includegraphics[width=0.6\linewidth]{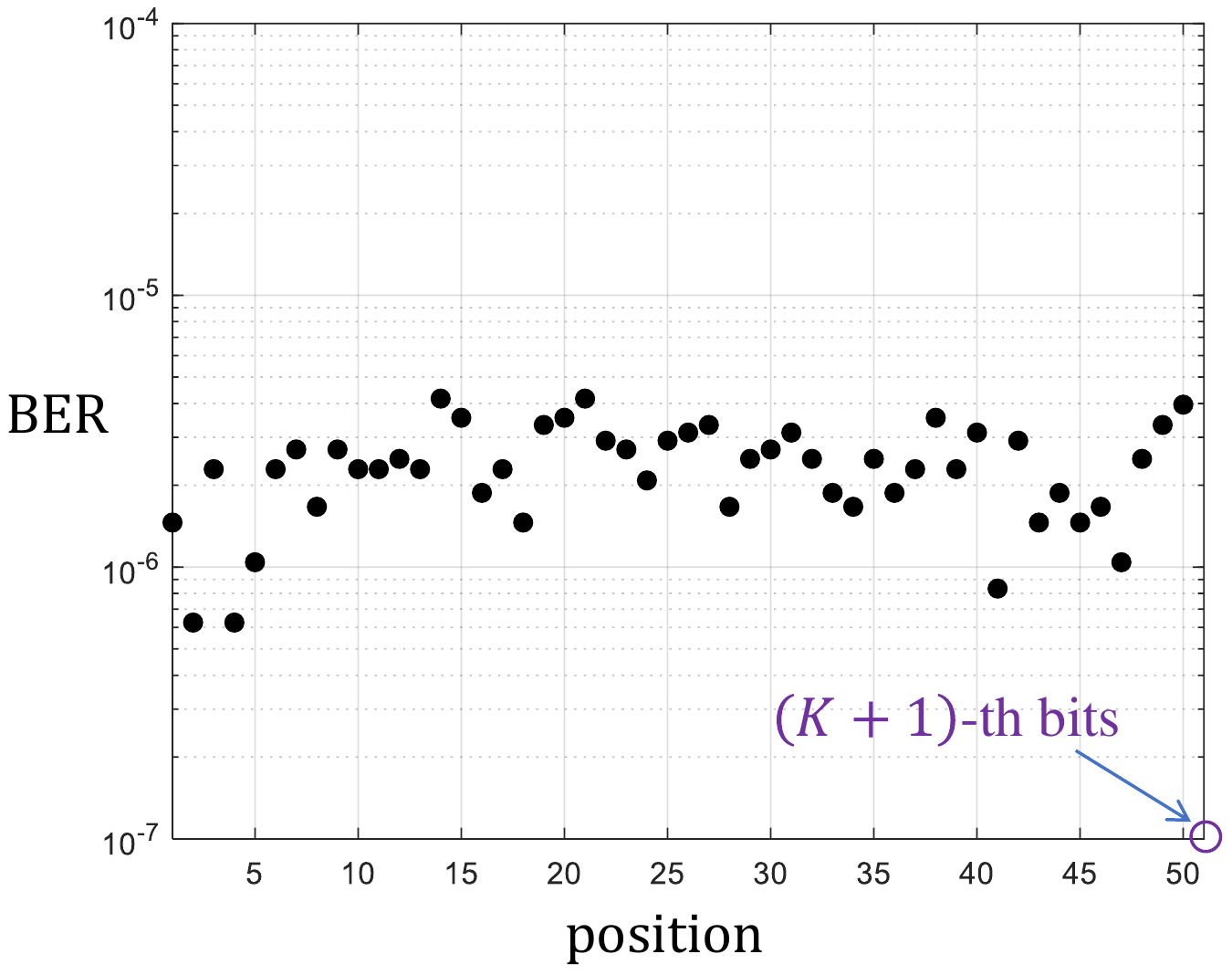}
\caption{BER versus the position of the bits.}
\label{fig:ber}
\end{figure}

{\it BER}: Considering the AWGN channel case with a feedforward SNR $\eta_r=0$ dB and a feedback SNR $\eta^\prime_r=20$ dB, the BER performance versus the bit position $k$ is plotted in Fig. \ref{fig:ber}.

\begin{figure}[t]
\centering
\includegraphics[width=0.6\linewidth]{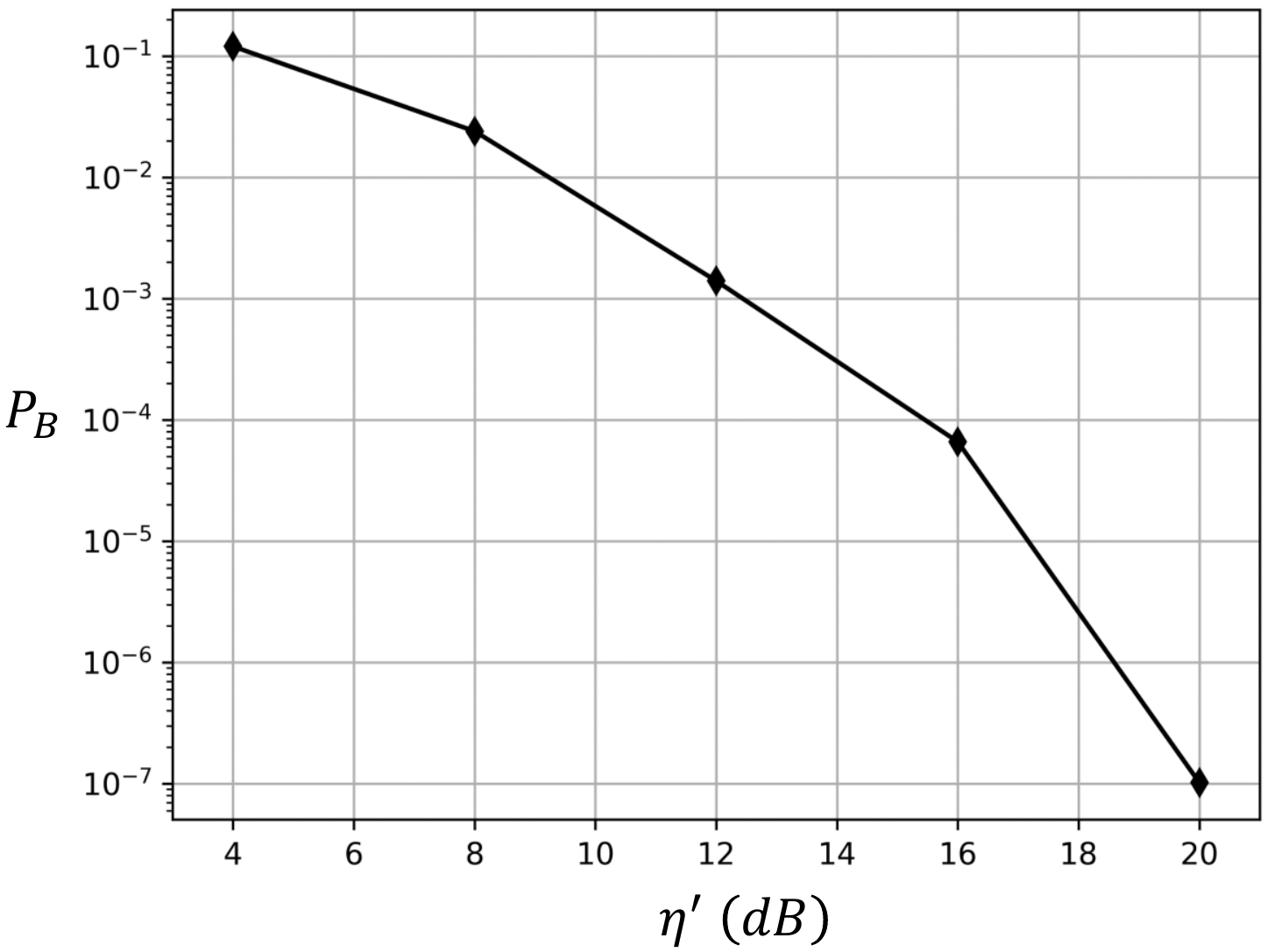}
\caption{The performance of AttentionCode versus the feedback channel SNR.}
\label{fig:noisyfeedback}
\end{figure}

{\it Impact of feedback noise}: The simulation results are shown in Fig. \ref{fig:noisyfeedback}, wherein we consider AWGN channels with a feedforward SNR $\eta=1$ dB and a feedback SNR $\eta^\prime\in[4,20]$ dB.
As can be seen, the BLER performance of AttentionCode deteriorates with the decrease in the feedback channel SNR. This is easy to understand since 1) when the feedback is very noisy, the useful information that can be exploited from the feedback is very limited; 2) when governed by noise, feedback can mislead the encoder, yielding error propagations across interactions.

In general, if we keep the SNRs of the feedforward and feedback channel to be similar (e.g., around $0$ dB), all existing feedback codes, including AttentionCode, cannot learn a structured encoding and decoding scheme that achieves low BLER. Designing good feedback codes in the presence of very noisy feedback is still an open problem that deserves further research.

\begin{figure}[t]
\centering
\includegraphics[width=0.6\linewidth]{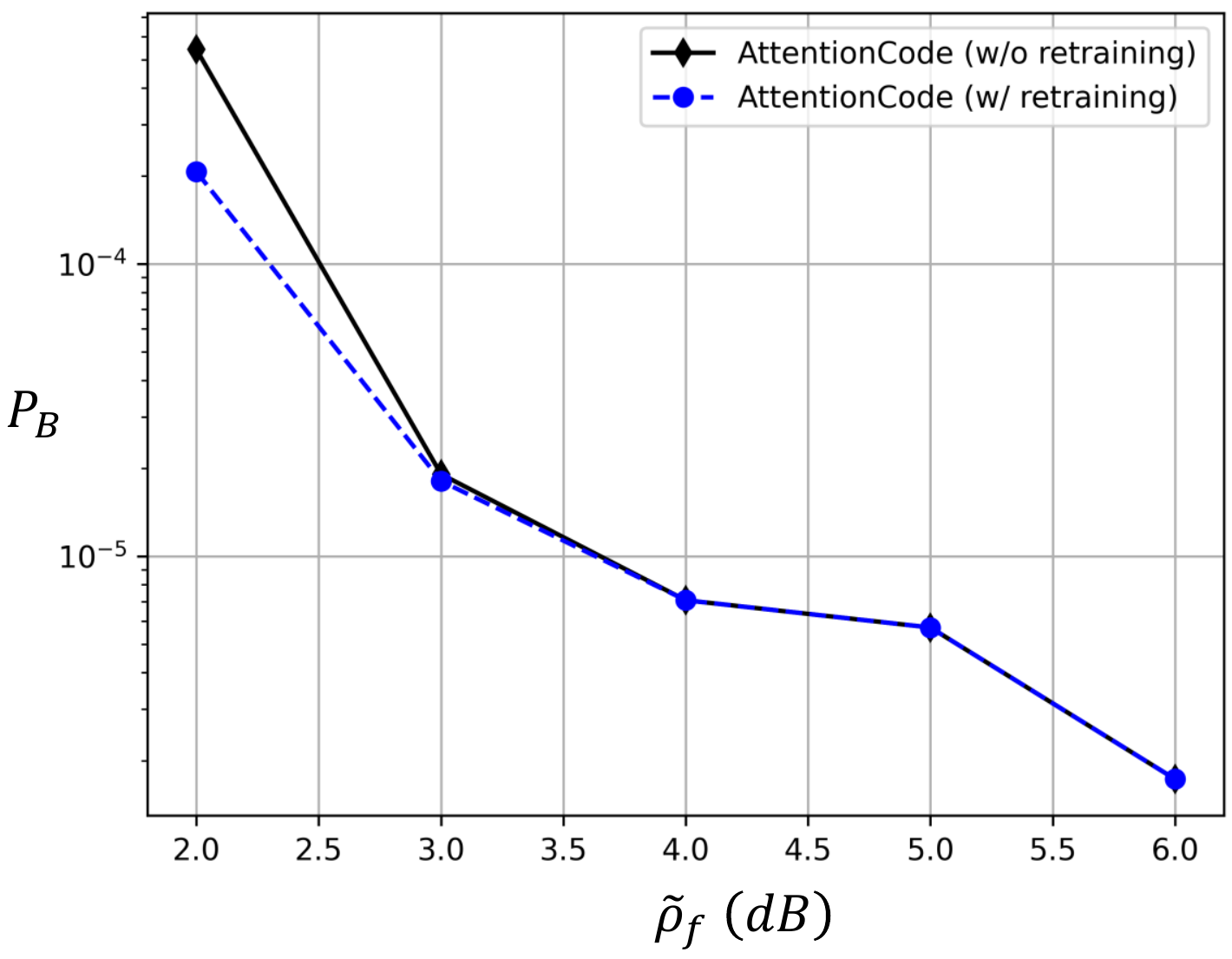}
\caption{Performance of AttentionCode (with and without retraining) when the test channels mismatch the training channel.}
\label{fig:robust}
\end{figure}

{\it Unseen channels}:  Consider the simulation in Fig.~\ref{fig:S3}(b). We extract the well-trained AttentionCode under an average feedforward SNR $\mathbb{E}[\eta_r]=25$ dB and an average feedback SNR $\mathbb{E}[\eta^\prime_r]=38.1$ dB. To evaluate AttentionCode in unseen channels, we vary the distribution of fading coefficients by changing $\widetilde{\rho}_f$. To ease exposition, we first define two channels:
\begin{itemize}
    \item {\it Training channel}: The channel that AttentionCode is trained on, i.e., the fading channel with parameters $\mathbb{E}[\eta_r]=25$ dB and $\mathbb{E}[\eta^\prime_r]=38.1$ dB.
    \item {\it Test channel}: The channel that the well-trained AttentionCode will be evaluated on.
\end{itemize}
First, we evaluate the performance of AttentionCode trained on the training channel on various test channels. The BLER versus $\widetilde{\rho}_f$ performance is shown in Fig. \ref{fig:robust} (black curve). As can be seen, despite the unseen channel distributions, AttentionCode still performs well, showing robustness to mismatched channel distributions.

Second, we evaluate the adaptability of AttentionCode by retraining it with a fixed number of 100 mini-batches. In each mini-batch, we randomly sample $2000$ bitstreams and channel realizations using the distribution of the test channel and retrain AttentionCode. As can be seen from Fig. \ref{fig:robust} (blue curve), with a limited number of retraining mini-batches, the retrained AttentionCode outperforms the original AttentionCode when the test channel is worse than the training channel. In contrast, when the test channel is better than the training channel, we do not observe performance gains by retraining.

\begin{figure}[t]
\centering
\includegraphics[width=0.6\linewidth]{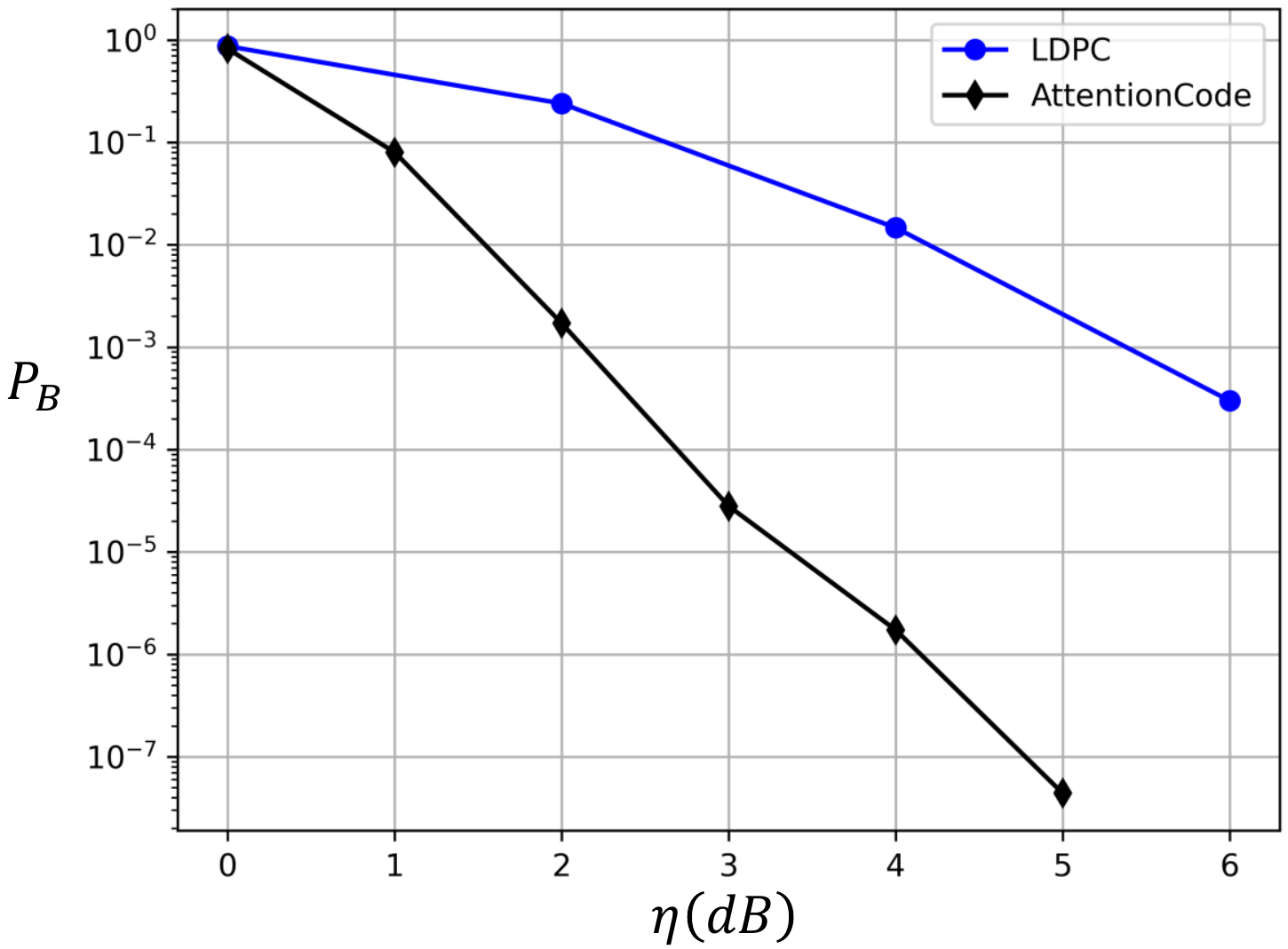}
\caption{Performance of AttentionCode and LDPC: $K=50$, $R=1/2$, AWGN channel, noiseless feedback.}
\label{fig:rate}
\end{figure}

{\it $1/2$ rate}: Fig. \ref{fig:rate} presents the performance of AttentionCode with a code rate $R=1/2$ and a block length $K=50$. AttentionCode outperforms LDPC by $3.7$ dB to achieve a BLER of $10^{-4}$.

{\it Neural network settings}: The detailed architecture and parameter setting of each layer are shown in Fig. \ref{fig:model}, where $d_S=4$ (Enc) or 3 (Dec), $k=1,2,...,K+1$, and $d_m=32$. The parameters are chosen based on extensive simulations and evaluations, and the experience reported in the literature.

\begin{figure}[t]
\centering
\includegraphics[width=0.6\linewidth]{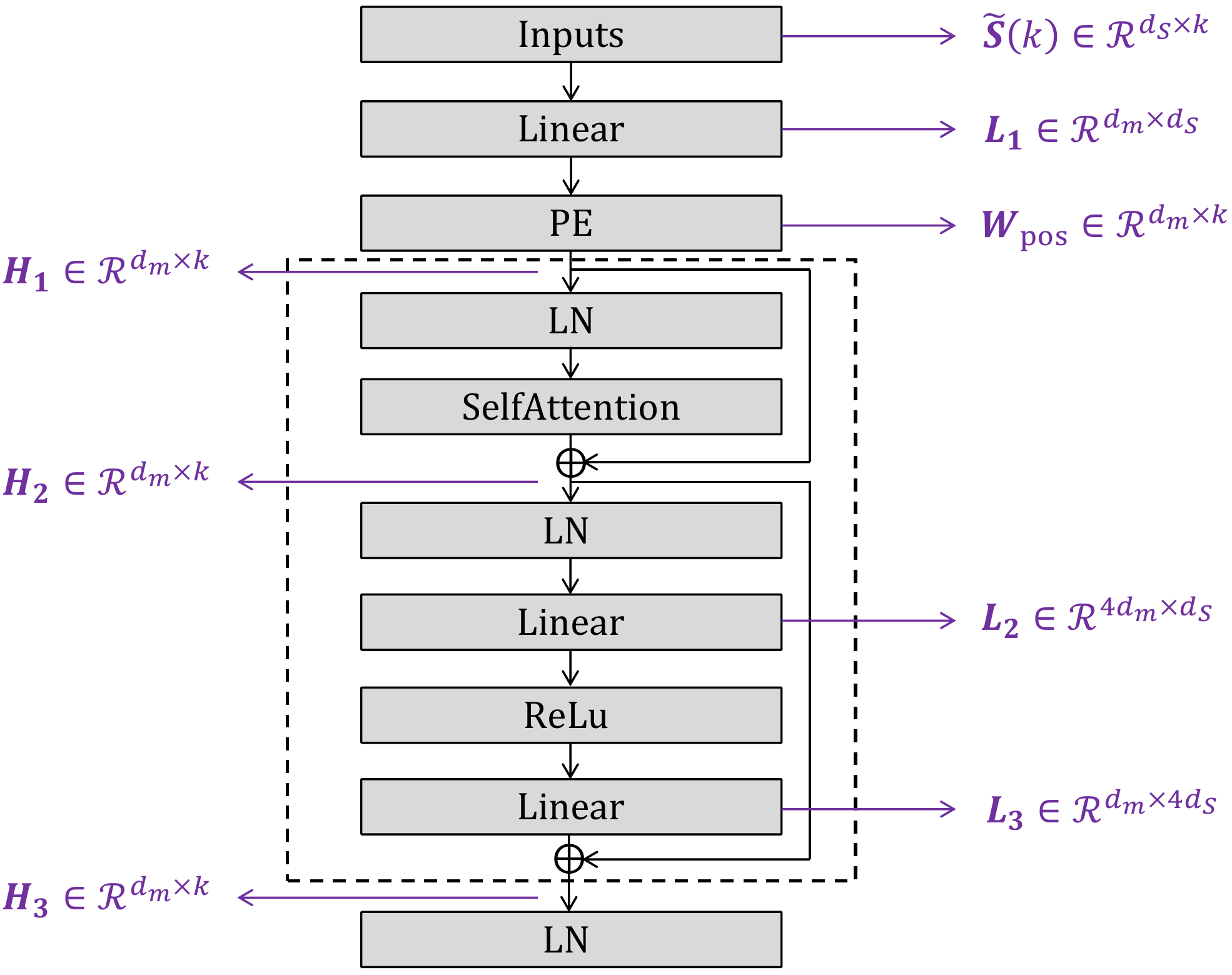}
\caption{Parameter setting for AttentionCode.}
\label{fig:model1}
\end{figure}

\begin{figure}[t]
\centering
\includegraphics[width=0.6\linewidth]{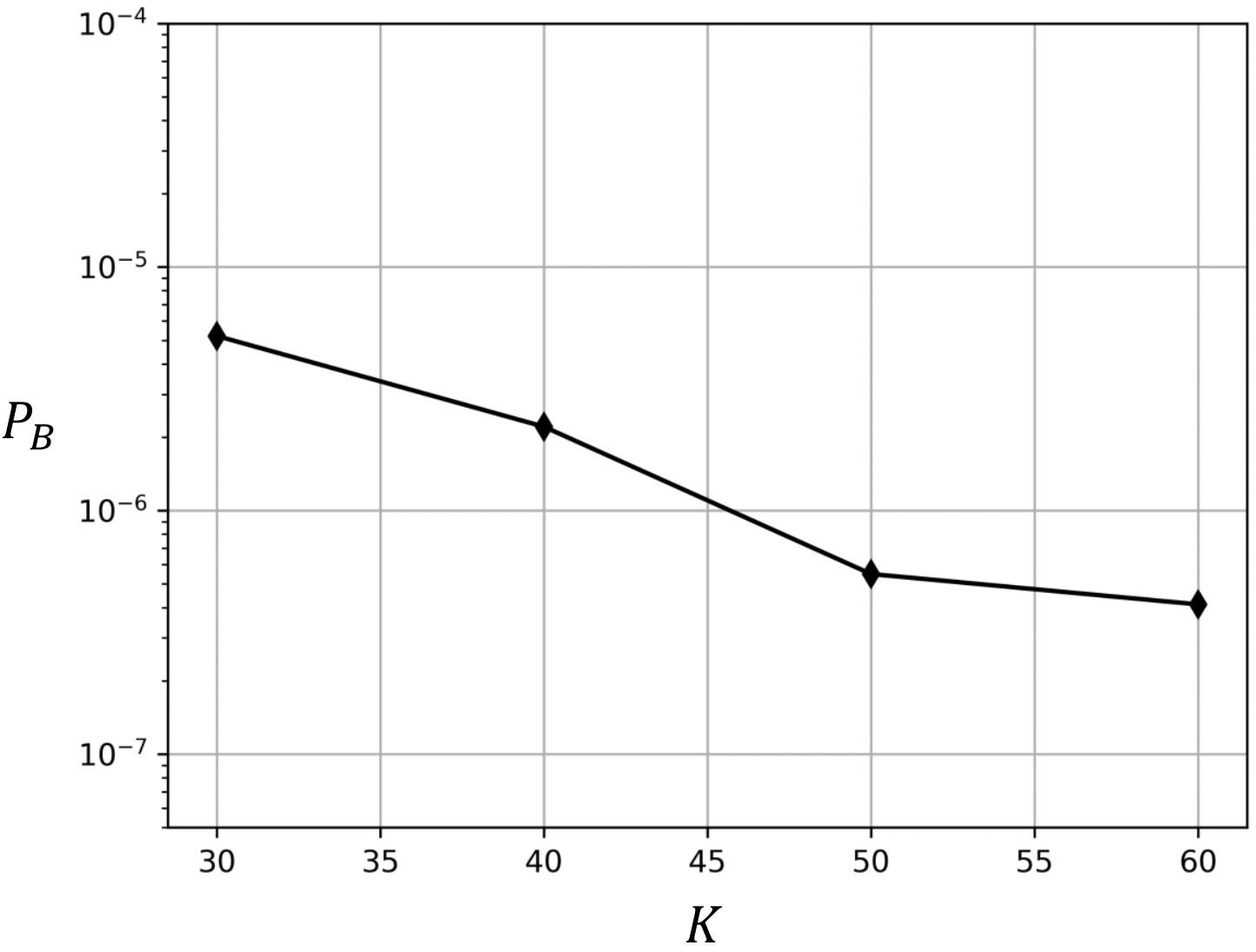}
\caption{BLER performance of AttentionCode versus the block length $K$.}
\label{fig:KK}
\end{figure}

{\it Block length}: 
Fig. \ref{fig:KK} presents additional simulations with different $K\in[30,60]$ to evaluate the BLER performance of AttentionCode. As can be seen, the BLER decreases with the increase in $K$, which is expected.

It is worth noting that AttentionCode cannot achieve very low BLER when $K$ is large (e.g. $K=100$). This is due to the increasing training complexity and the small batch size constrained by the limited GPU resources. To address this problem, an efficient solution is grouping the bits into small blocks to generate parity symbols, thereby reducing the input sequence length (hence the complexity) to AttentionNet.

\bibliographystyle{IEEEtran}
\bibliography{References}

%
%

\end{document}